 \documentclass[twocolumn]{aastex6}
\newcommand{\oiii}{[O{\,\sc iii}]\,\,}
\newcommand{\oii}{[O{\,\sc ii}]\,\,}
\newcommand{\neiii}{[Ne{\,\sc iii}]\,\,}
\newcommand{\sii}{[S{\,\sc ii}]\,\,}

\fullcollaborationName{The Friends of AASTeX Collaboration}

\begin{document}


%\title{An Example Article using \aastex v6.0}
\title{Ionized gas outflows in infrared-bright dust-obscured galaxies selected with WISE and SDSS}

%% Use \author, \affil, plus the \and command to format author and affiliation 
%% information.  If done correctly the peer review system will be able to
%% automatically put the author and affiliation information from the manuscript
%% and save the corresponding author the trouble of entering it by hand.
%%
%% The \affil should be used to document primary affiliations and the
%% \altaffil should be used for secondary affiliations, titles, or email.

%% Authors with the same affiliation can be grouped in a single

\author{Yoshiki Toba 		\altaffilmark{1,2},
		Hyun-Jin Bae 		\altaffilmark{3,4,5},
		Tohru Nagao 		\altaffilmark{2},
		Jong-Hak Woo		\altaffilmark{3},
		Wei-Hao Wang		\altaffilmark{1},
		Alexander Y. Wagner \altaffilmark{6,7},
		Ai-Lei Sun 			\altaffilmark{1,8},
		Yu-Yen Chang		\altaffilmark{1}
		}	  
\affil{}  			  
  \altaffiltext{1}{Academia Sinica Institute of Astronomy and Astrophysics, PO Box 23-141, Taipei 10617, Taiwan}
  \email{toba@asiaa.sinica.edu.tw}
  \altaffiltext{2}{Research Center for Space and Cosmic Evolution, Ehime University, Bunkyo-cho, Matsuyama, Ehime 790-8577, Japan} 
  \altaffiltext{3}{Astronomy Program, Department of Physics and Astronomy, Seoul National University, Seoul 151-742, Korea}
  \altaffiltext{4}{Department of Astronomy and Center for Galaxy Evolution Research, Yonsei University, Seoul 120-749, Korea}   
  \altaffiltext{5}{Biomedical engineering research center, Asan Medical Center, Seoul 05505, Korea}  
  \altaffiltext{6}{Center for Computational Sciences, University of Tsukuba, 1-1-1 Tennodai, Tsukuba, Ibaraki 305-8577, Japan}
  \altaffiltext{7}{Institut d'Astrophysique de Paris 98 bis bd Arago, F-75014 Paris, France}
  \altaffiltext{8}{Department of Physics and Astronomy, Bloomberg Center, Johns Hopkins University, Baltimore, MD 21218, USA}

%\author{Greg J. Schwarz\altaffilmark{1,2} and August Muench\altaffilmark{1}}
%\affil{American Astronomical Society \\
%2000 Florida Ave., NW, Suite 300 \\
%Washington, DC 20009-1231, USA}

%\author{Butler Burton\altaffilmark{3}}
%\affil{National Radio Astronomy Observatory}

%\affil{TeXnology Inc}

%\author{Julie Steffen\altaffilmark{4}}
%\affil{American Astronomical Society \\
%2000 Florida Ave., NW, Suite 300 \\
%Washington, DC 20009-1231, USA}

%% Use the \and command so offset the last author.
%\and

%\author{Jeff Lewandowski\altaffilmark{5}}
%\affil{IOP Publishing, Washington, DC 20005}


%\altaffiltext{1}{AAS Journals Data Scientist}
%\altaffiltext{2}{greg.schwarz@aas.org}
%\altaffiltext{3}{AAS Journals Associate Editor-in-Chief}
%\altaffiltext{4}{AAS Director of Publishing}
%\altaffiltext{5}{IOP Senior Publisher for the AAS Journals}

\begin{abstract}
We present the ionized gas properties of infrared (IR)-bright dust-obscured galaxies (DOGs) that show an extreme optical/IR color, $(i - [22])_{\rm AB} > 7.0$, selected with the Sloan Digital Sky Survey (SDSS) and Wide-field Infrared Survey Explorer ({\it WISE}).
For 36 IR-bright DOGs that show [O{\,\sc iii}]$\lambda$5007 emission in the SDSS spectra, we performed a detailed spectral analysis to investigate their ionized gas properties.
In particular, we measured the velocity offset (the velocity with respect to the systemic velocity measured from the stellar absorption lines) and the velocity dispersion of the \oiii line.
We found that the derived velocity offset and dispersion of most IR-bright DOGs are larger than those of  Seyfert 2 galaxies (Sy2s) at $z < 0.3$, meaning that the IR-bright DOGs show relatively strong outflows compared to Sy2s.
This can be explained by the difference of IR luminosity contributed from active galactic nucleus, $L_{\rm IR}$ (AGN), because we found that (i) $L_{\rm IR}$ (AGN) correlates with the velocity offset and dispersion  of \oiii and (ii) our IR-bright DOGs sample has larger $L_{\rm IR}$ (AGN) than Sy2s.
Nevertheless, the fact that about 75\% IR-bright DOGs have a large ($>$ 300 km s$^{-1}$) velocity dispersion, which is a larger fraction compared to other AGN populations, suggests that IR-bright DOGs are  good laboratories to investigate AGN feedback.
The velocity offset and dispersion of \oiii and [Ne{\,\sc iii}]$\lambda$3869 are larger than those of [O{\,\sc ii}]$\lambda$3727, which indicates that the highly ionized gas tends to show more stronger outflows.
\end{abstract}




%%%%%%%%%%%%%%%%%%%%%%
%	 INTRODUCTION
%%%%%%%%%%%%%%%%%%%%%%
\section{Introduction}

It has been well-known that the mass of the super massive black hole (SMBH) tightly correlates with the properties of its host galaxy such as spheroid component mass and stellar velocity dispersion, which suggests that SMBHs and galaxies coevolve \citep[so-called ``co-evolution'': e.g.,][]{Magorrian,Marconi,Kormendy,McConnell,Sun,Woo_13}. 
Observations at various wavelengths have indicated that radiation, winds, and jets from an active galactic nucleus (AGN) can interact with the interstellar medium, and this can lead to the ejection or heating of gas.
Therefore, AGN feedback has been increasingly considered as a key component to understand the galaxy formation and evolution \cite[e.g.,][and references therein]{Fabian}, which is also supported by hydrodynamical simulations \cite[e.g.,][]{Wagner_11,Faucher-Giguere,Wagner_13,Bieri}.
These powerful outflows resulting from feedback caused by the AGN regulate star formation (SF) and even AGN activity, and could control co-evolution of galaxies and SMBHs \citep[e.g.,][]{Di,Cano-Diaz,King}.
Measuring the kinematics of multiphase gas is one of the useful ways to investigate gas outflows in AGNs. 
In particular, the velocity offset of the [O{\,\sc iii}]$\,\lambda$5007\AA \, narrow emission and its velocity dispersion are good tracers for probing AGN-driven outflows.
Many works have reported strong [O{\,\sc iii}] outflows in AGNs \cite[e.g.,][]{Zamanov,Aoki,Bian,Boroson, Komossa,Crenshaw,Greene,Villar,Rodriguez,Liu_a,Liu_b,Mullaney,Zakamska_14,Sun_17} and investigated their statistical properties \citep[e.g.,][]{Wang_a,Bae,Woo,Woo_17}.
The advent of the integral field unit (IFU) enables us to investigate AGN feedback providing spatial information of AGN outflows from local Universe \citep[e.g.,][]{Barbosa,Harrison,McElroy,Karouzos_16b,Karouzos_16b,Bae_17} to high-z Universe \citep[e.g.,][]{Alexander,Brusa,Carniani}.

In this paper, we present the ionized gas properties of IR-bright dust-obscured galaxies \citep[DOGs:][]{Dey, Toba_15} that show an extreme optical and IR color, i.e., their flux densities in the mid-IR (MIR) regime are about 1000 times brighter than those in the optical regime, indicating that these objects are undergoing strong AGN and/or SF activity behind the large amount of dust.
We have performed IR-bright DOGs search and investigated their statistical and physical properties such as IR luminosity function \citep{Toba_15}, auto-correlation function \citep{Toba_17a}, and stellar mass and star-formation rate relation \citep{Toba_17b}.
The IR luminosity of most of the IR-bright DOGs exceeds $10^{12} L_{\sun}$ or even $10^{13} L_{\sun}$, which are termed ultraluminous infrared galaxies \citep[ULIRGs:][]{Sanders} and hyperliminous infrared galaxies \citep[HyLIRGs:][]{Rowan-Robinson}, respectively.
In the context of major merger scenario, the gas accreting onto the nucleus triggers the AGN activity due to the merger process, and enormous energy originated from the AGNs then significantly affects SF activity in the host galaxies \citep[e.g.,][]{Hopkins_06,Hopkins_08}.
Since IR-bright DOGs may correspond to a maximum phase of AGN activity behind large amount of dust \citep[e.g.,][]{Narayanan}, they are expected to be a good laboratory to investigate the AGN feedback phenomenon.
Note that observations of molecular and atomic gas are quite useful to investigate the kinematics and energetics of outflowing gas \citep[e.g.,][]{Cicone}.
However, these investigations often require follow-up observations with radio telescopes and the sample size is limited due to the low efficiency of these observations.
In order to investigate the statistical aspect of outflowing gas in IR-bright DOGs, we focus on ionized gas.

This paper is organized as follows. 
We describes the sample selection and spectral analysis in Section \ref{DA}.
The resultant outflow properties of \oiii is presented in Section \ref{Rs}.
In Section \ref{Dis}, we discuss the dependence of the \oiii outflow properties on physical properties such as IR luminosity. 
We also discuss the energetics of AGN outflow in our sample and present the outflow properties of other emission lines. 
We summarize in Section \ref{Sum}.
Throughout this paper, the adopted cosmology is a flat universe with $H_0$ = 70 km s$^{-1}$ Mpc$^{-1}$, $\Omega_M$ = 0.3, and $\Omega_{\Lambda}$ = 0.7. Unless otherwise noted, all magnitudes refer on the AB system. and we adopt vacuum wavelengths for the analysis.\\

%%%%%%%%%%%%%%%%%%%%%%
% DATA and ANALYSIS
%%%%%%%%%%%%%%%%%%%%%%
\section{Data and analysis}
\label{DA}
   \begin{figure}
   \centering
   \includegraphics[width=0.45\textwidth]{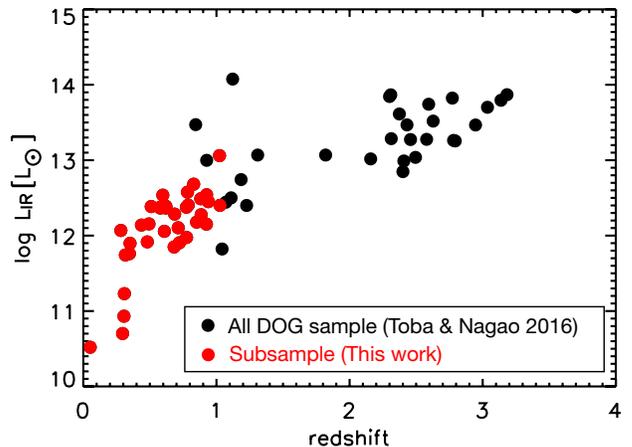}
   \caption{IR luminosity as a function of redshift for our IR-bright DOG sample, discovered by \cite{Toba_16}. Red circles show objects that are used for the spectral analysis in this work.}
   \label{LIR}
   \end{figure}
      \begin{figure*}[t]
   \centering
   \includegraphics[width=0.95\textwidth]{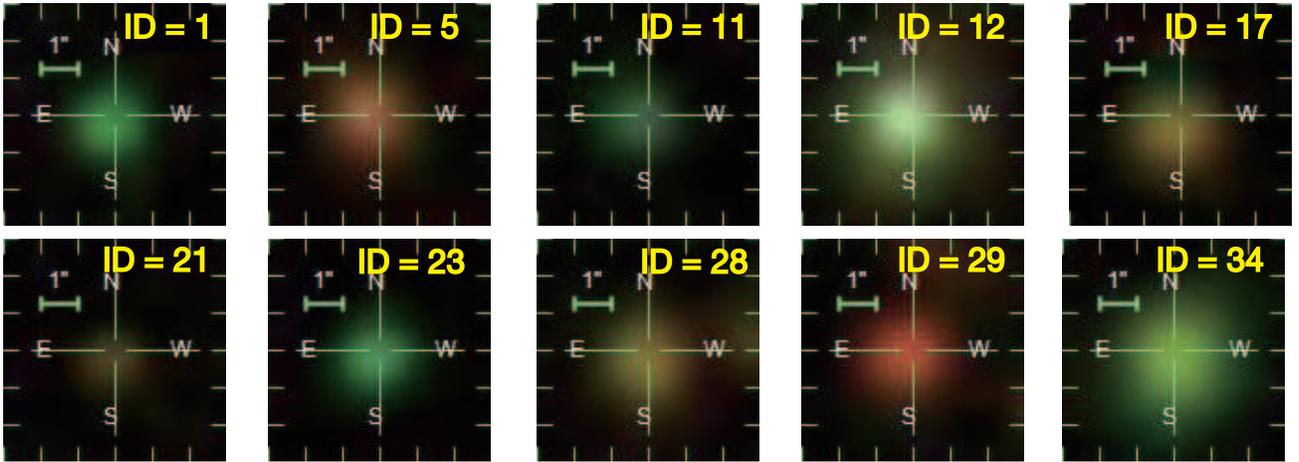}
   \caption{Example of the SDSS $gri$-composite  images of our IR-bright DOG sample.}
   \label{images}
   \end{figure*} 
The DOG Sample for the spectral analysis was selected from a IR-bright DOG sample in \cite{Toba_16}.
They selected 67 IR-bright DOGs with $(i - [22])_{AB} > 7.0$ and flux density at 22 $\micron$ $>$ 3.8 mJy from the Sloan Digital Sky Survey (SDSS) spectroscopic catalog \citep{York,Alam} and {\it Wide-field Infrared Survey Explorer} ({\it WISE}) ALLWISE catalog \citep{Wright,Cutri}.
Among them, we narrowed down to 36 objects with 0.05 $< z <$ 1.02 that clearly have \oiii in their SDSS spectra \footnote{Within our sample, the spectra of SDSS1010+3775 (ID=12), SDSS1248+4242 (ID=21), SDSS1407+3601 (ID=26), and SDSS1513+1451 (ID=30) have also reported in \cite{Ross}.They also mentioned that SDSS1010+3775 has unusually broad \oiii with non-Gaussian structure.}.
Figure \ref{LIR} shows IR luminosity, $L_{\rm IR}$ (8--1000 $\micron$), as a function of redshift for all IR-bright DOG sample and this subsample.
The IR luminosities of the 36 IR-bright DOG sample are $\log L_{\rm IR}$ [$L_{\sun}$] = 10.5 -- 13.1, and 25/36 ($\sim$ 69 \%) objects are classified as ULIRGs/HyLIRGs (see Table \ref{table}).
Recently some authors have discovered many (obscured) ULIRGs/HyLIRGs based on the SDSS and {\it WISE} data and reported powerful ionized outflows seen in their spectra \citep{Ross,Zakamska_16,Bischetti,Hamann,Zhang_S}, although they basically focus on high-z ($z > 2$) objects.

The SDSS spectra for all 36 IR-bright DOGs are shown in Figure \ref{spectra}--\ref{spectra4} (see Appendix \ref{spectra_all}).
The mean full width at half maximum (FWHM) of their H$\beta$ line is about 503 km s$^{-1}$.
When we adopt 1000 km s$^{-1}$ as a threshold to discriminate between type 1 and 2 AGNs \citep[e.g.,][]{Yuan}, 4/36 objects can be classified as a type 1 AGN, meaning that most objects in our DOG sample are type 2 AGNs (see Table \ref{table}).
One prominent feature in these spectra is that they often show broad asymmetric profiles of \oiii lines, which could indicate some IR-bright DOGs are blowing out ionized gas.
In order to characterize this \oiii outflow quantitatively, we need to perform a detailed spectral fitting  for each spectrum.
Since most objects in our sample are type 2 AGNs, the stellar continuum can be seen, which enables us to measure systemic velocity determined by stellar fitting and to estimate velocity offset with respect to the systemic velocity (see Section \ref{Spfit}). 

Therefore, we conducted the spectral analysis for 36 IR-bright DOGs to quantify the \oiii outflow, in the same manner as \cite{Bae} (see also references therein).
First, we subtracted the stellar continuum by using the templates of simple stellar population models \citep[MILES;][]{Sanchez-Blazquez}, and we measured the velocity of the luminosity-weighted stellar component of the host galaxy (systemic velocity) based on the best-fit model.
The fitting is based on the Penalized Pixel-Fitting method \citep[{\tt pPXF};][]{Cappellari}.
The typical error of the measured systemic velocity is 52.6 km s$^{-1}$.
For the starlight-subtracted spectra, we fitted the H$\beta$ and \oiii doublet ([O{\,\sc iii}]$\lambda$4959, 5007) with a single- and double-Gaussian function separately using {\tt MPFIT}, an IDL $\chi^2$-minimization routine \citep{Markwardt}. 
We assume that the H$\beta$ and \oiii doublet have independent kinematics, while the \oiii lines (4959\AA \,\,and 5007\AA) have the same velocity and velocity dispersion to each other. 
If the peak amplitude of broad component between the two Gaussian profiles is larger than the continuum noise (i.e., the amplitude-to-noise ratio is larger than 2), we adopted the fitting results with a double-Gaussian function. 
Otherwise, we adopted the result with a single Gaussian.
Note that we visually checked whether the stellar continuum is reproduced well by the best-fit stellar fitting.
We confirmed that 10/36 objects are well-fitted by the stellar template.
For the remaining 26 objects, we alternatively utilized the narrow component of the H$\beta$ line as a proxy of the systemic velocity.  \\

%%%%%%%%%%%%%%%%%%%%%%
%	   RESULTS
%%%%%%%%%%%%%%%%%%%%%%
\section{Results}
\label{Rs}
   \begin{figure*}
   \centering
   \includegraphics[width=0.95\textwidth]{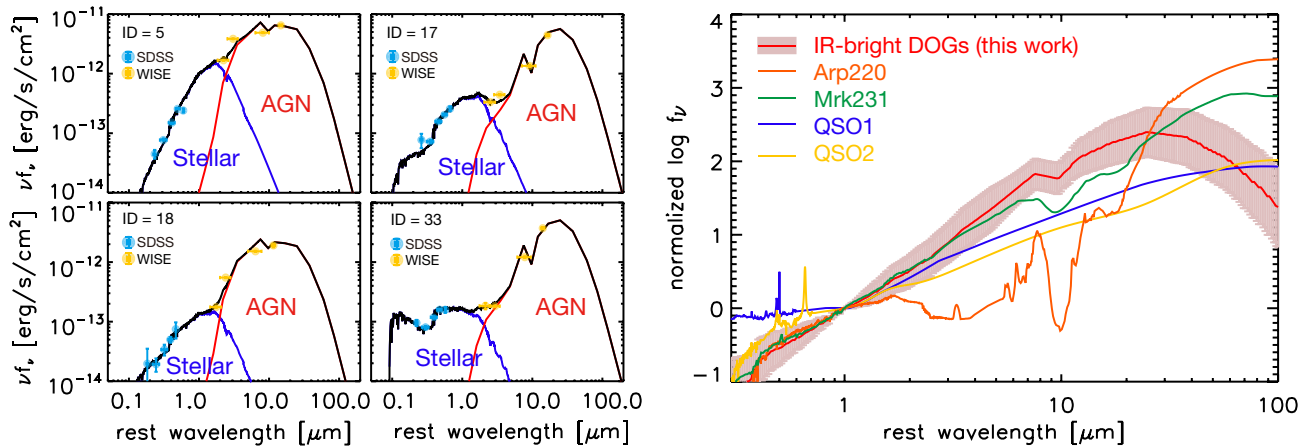}
   \caption{(Left) Examples of the SED fitting for the IR-bright DOG sample. The cyan and yellow circles represent the data from the SDSS and WISE, respectively. The contribution from the stellar and AGN components to the total SEDs are shown in blue and red lines, respectively. The black solid line represents the resultant (the combination of the stellar and AGN) SEDs. (Right) Comparison of composite spectrum of IR-bright DOGs (red line with shaded region) with other SED templates of local ULIRGs/AGNs presented by \cite{Polletta} Each SED is normalized by the flux density at 1 $\micron$.}
   \label{SED}
   \end{figure*}  
%------------------------
%    Spectral fitting
%------------------------
\subsection{Spectral fitting}
\label{Spfit}
Using the best fit with single- or double- Gaussian components, we measured the velocity offset ($v_{\rm line}$) and velocity dispersion ($\sigma_{\rm line}$) in the same manner as \cite{Woo};
\begin{eqnarray}
v_{\rm line} \, (\lambda)  & = & \frac{(\lambda_0 - \lambda_{\rm rest})c}{\lambda_{\rm rest}}  - v_{\rm sys} \, (\lambda), \\
\sigma_{\rm line} \, (\lambda) & = & \sqrt{\frac{\int \lambda^2 f(\lambda) \,{\rm d}\lambda}{\int f(\lambda) \,{\rm d}\lambda} - \lambda_0^2},
\end{eqnarray}
were  $\lambda_{\rm rest}$ is the rest-frame line center of a line ($\lambda_{\rm rest}$  = 5008.24 \AA \,\, for [O{\,\sc iii}]), and $c$ is the speed of light, while $v_{\rm sys} \, (\lambda)$ is the systemic velocity measured by the fitting with a stellar component or a narrow component of H$\beta$ (see Section \ref{DA}).
$f(\lambda)$ is the flux density at each wavelength and $\lambda_0$ is the first moment of the line profile (flux-weighted center),
\begin{equation}
\lambda_0   =  \frac{\int \lambda f(\lambda) \,{\rm d}\lambda}{\int f(\lambda) \,{\rm d}\lambda}.
\end{equation}
The measured velocity dispersions were corrected for the wavelength-dependent instrumental resolution of the SDSS.
The measurement errors was estimated from a Monte Carlo realization; we adopted 1$\sigma$ dispersion of each value by measuring them 100 times for spectra with randomly adding the noise \citep[see][in detail]{Woo}.

The resultant velocity offset ($v_{\rm [OIII]}$) and dispersion ($\sigma_{\rm [OIII]}$) for \oiii line of our DOG sample are tabulated in Table \ref{table}.
We found that 29/36 objects show a broad wing of [O{\,\sc iii}] (we labeled them as $w_{\rm [OIII]}$ = 1; see Table \ref{table}), and thus we fit them with double Gaussian.
For the remaining 7 objects ($w_{\rm [OIII]}$ = 0), we fit them with single Gaussian. 
Figure \ref{images} shows some examples of the SDSS composite images made by $g$, $r$, and $i$ images.  
Some DOGs show a green or red color since strong \oiii line fall in the $r$- or $i$-band, depending on the redshift.\\

Hereafter we compare outflow properties of our IR-bright DOG sample with those of local Seyfert 2 galaxies (Sy2s).
In order to ensure a fair comparison, we only focus on 36--4 = 32 IR-bright DOGs that are classified as type 2 AGNs unless otherwise noted.
Note that among 32 DOGs, 12 objects have very large uncertainties of velocity offset ($\delta_{v_{\rm [OIIII]}}$), i.e., $\delta_{v_{\rm [OIIII]}} > v_{\rm [OIII]}$ although all objects have $\delta_{\sigma_{\rm [OIIII]}} < \sigma_{\rm [OIII]}$.
We exclude them and focus on 32--12 = 20 DOGs when arguing about the velocity offset.
We found that 24/32 ($\sim$ 75 \%) IR-bright DOGs have a large ($>$ 300 km s$^{-1}$) velocity dispersion, which is larger than that of local Sy2s at $z < 0.3$ \citep{Woo} who reported that only 3.58 \% of Sy2 sample show $\sigma_{\rm [OIII]}$ $>$ 300 km s$^{-1}$. 
Also, 19/20 ($\sim$95 \%) DOGs have $|v_{\rm [OIII]}| >$ 50 km s$^{-1}$, that is larger than those ($\sim$ 50\%) of local (narrow line) Seyfert 1 and 2 galaxies \citep[e.g.,][]{Komossa,Zhang_K,Bae}.
Since the velocity offset and (particularly) velocity dispersion is expected to be due to the ionized gas outflow, this large outflow fraction could indicate that IR-bright DOGs are likely to be a good laboratory to investigate AGN feedback phenomenon (see also Section \ref{sVVD}).

%------------------------
% Luminosity correlations
%------------------------
\subsection{Relation between \oiii luminosity and IR luminosity}
\label{LL}

   \begin{figure*}
   \centering
   \includegraphics[width=0.8\textwidth]{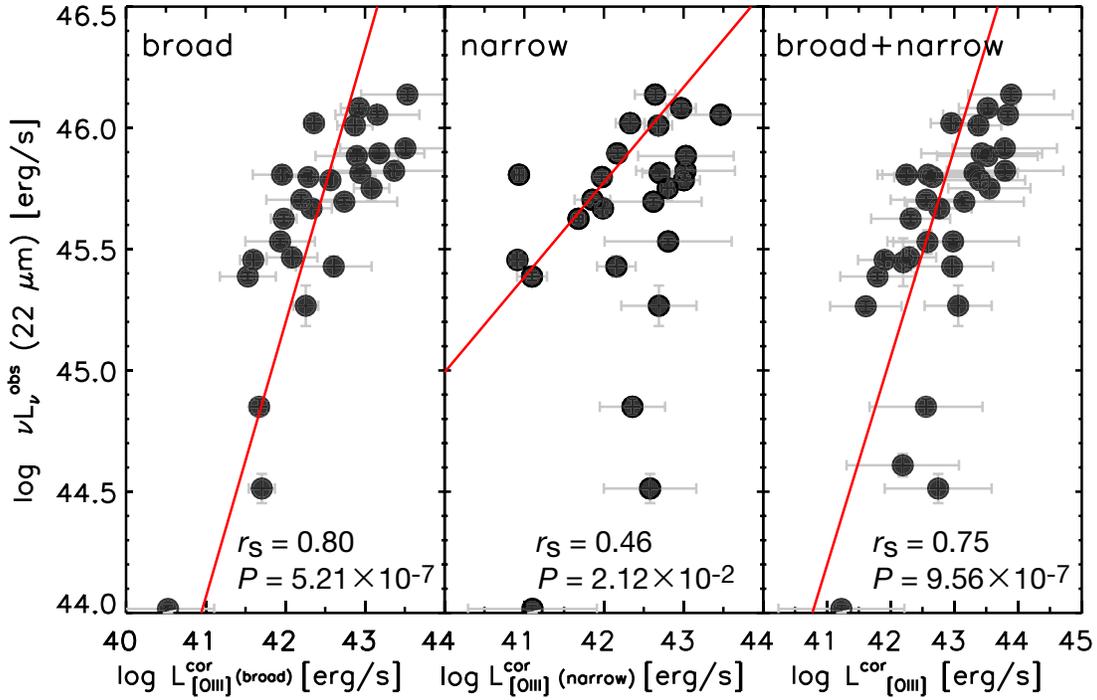}
   \caption{The 22 $\micron$ luminosity at observed frame as a function of \oiii luminosity of broad component (left), \oiii luminosity of narrow component (middle), and \oiii luminosity (right). The red line shows the best-fit linear function. The Spearman rank correlation coefficients ($r_{\rm s}$) with null hypothesis probabilities ($P$) for each relation are noted at the bottom right of each panel.}
   \label{LOIII_L22}
   \end{figure*}   
Here we estimated the extinction-corrected \oiii luminosity using the following formula \citep[see][]{Calzetti_94,Dominguez};
\begin{equation}
L_{\rm [OIII]}^{\rm cor} = L_{\rm [OIII]}^{\rm obs} 10^{0.4 k_{\rm [OIII]}\, E(B-V)},
\end{equation}
where $L_{\rm [OIII]}^{\rm obs}$ is the observed \oiii luminosity, $k_{\rm [OIII]}$ is the extinction value  at $\lambda = 5008.24$ \AA \,\,provided by \cite{Calzetti}, and $E(B - V)$ is the color excess.
We note that $E (B-V)$ was estimated based on the spectral energy distribution (SED) fitting with a code; SEd Analysis using BAyesian Statistics ({\tt SEABASs}: \citealt{Rovilos}).
This fitting code provides up to three-component fitting (AGN, SF, and stellar component) based on the maximum likelihood method \citep[see][in detail]{Rovilos,Toba_16}.
Among three components fitting, $E (B-V)$ was determined by the stellar component fitting with a library of synthetic stellar templates from \cite{Bruzual} stellar population models reddened using a \cite{Calzetti} dust extinction law.
We used 9 photometric data ($u$, $g$, $r$, $i$, and $z$, and 3.4, 4.6, 12, and 22 $\micron$, obtained from the SDSS and {\it WISE}, respectively) for the SED fitting.
We note that all DOGs in our sample were detected in all 9 bands.
The typical value of $E (B-V)$ is 0.70.
We also calculated the 22 $\micron$ luminosity at the observed frame, $\nu L^{\rm obs}_{\nu}$ (22 $\micron$), from the observed flux multiplied by $4\pi d_{\rm L}^2$ for each DOG, where $d_{\rm L}$ is the luminosity distance.
IR-bright DOGs tend to have flat SED at the MIR regime \citep[see ][]{Toba_16,Toba_17b} and we found that $\nu L^{\rm obs}_{\nu}$ (22 $\micron$) is perfectly correlated with IR luminosity \citep{Toba_17b}.
In addition, some authors claimed that IR luminosity of AGNs are correlated with \oiii luminosity \cite[e.g.,][]{Goto}, suggesting that 22 $\micron$ luminosity at observed frame correlates with \oiii luminosity.

    \begin{figure*}
   \centering
   \includegraphics[width=0.75\textwidth]{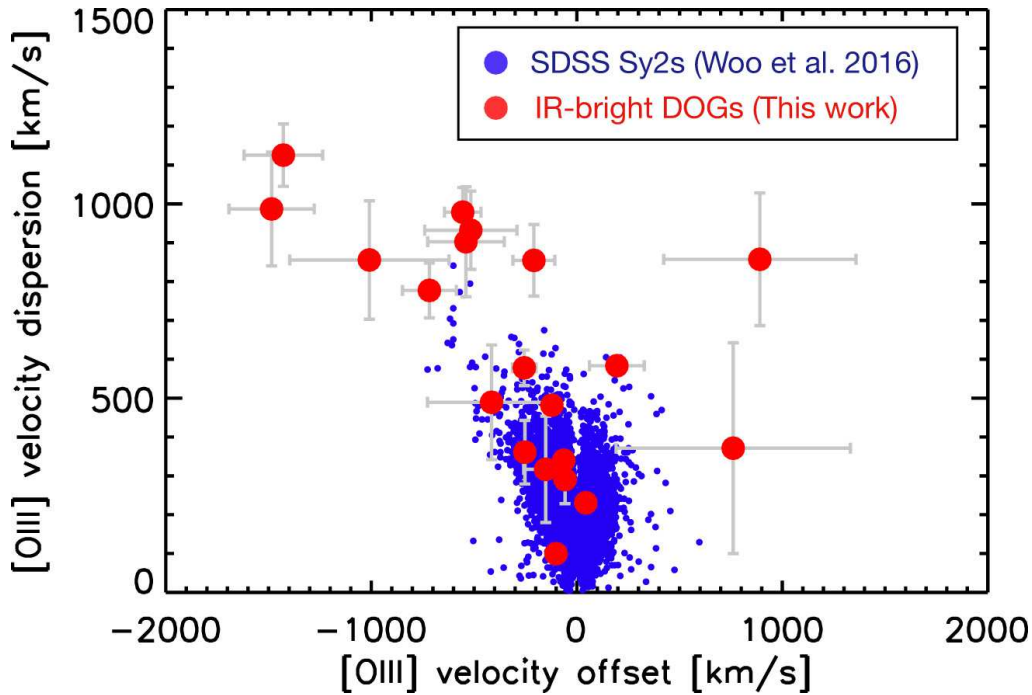}
   \caption{\oiii velocity offset -- velocity dispersion (VVD) diagram for IR-bright DOGs (red circle) and SDSS Seyfert 2 galaxies (blue circle) derived by \cite{Woo}.}
   \label{VVD}
   \end{figure*}  
   
Figure \ref{SED} shows examples of the SED fitting in which the data are well-fitted by {\tt SEABASs} \citep[see also][]{Toba_16}.
Their composite spectrum normalized by the flux density at 1 $\micron$ is also shown in this Figure.
Some SED templates of local ULIRGs and AGNs presented by \cite{Polletta} are also plotted.
Compared with these templates, our IR-bright DOG sample shows a steep SED in the near-IR (NIR) and MIR regions that could be originated from hot dust heated by strong AGN radiations. 

Figure \ref{LOIII_L22} shows the relation between \oiii luminosity and 22 $\micron$ luminosity at observed-frame.
As many authors have reported that \oiii luminosity are well-correlated with MIR luminosity \cite[e.g., ][]
{Toba_14,Yuan,Sun_17} for SDSS galaxies/AGNs, we confirmed that \oiii luminosity correlates with 22 $\micron$ (but at observed-frame) luminosity for our DOG sample, which is useful to infer the expected \oiii luminosity for IR-bright DOG from 22 $\micron$ flux density without considering the $k$-correction.
The relations between the \oiii luminosity for each of the broad and narrow component and $\nu L^{\rm obs}_{\nu}$ (22 $\micron$) are also shown in Figure \ref{LOIII_L22}.
Note that if an object does not have broad \oiii wing (see Section \ref{DA}), we derive extinction corrected \oiii luminosity based on result with single Gaussian fitting, and use them as $L_{\rm {[OIII]}}^{\rm cor}$ (broad+narrow).
In other words, \oiii luminosity of broad and narrow component in left and middle panel of Figure \ref{LOIII_L22} are derived only from objects with broad wing.
We fitted each relation with linear regression lines using a IDL routine, {\tt MPFITEXY}, that takes into account errors in both variables.  
The Spearman rank correlation coefficients ($r_{\rm s}$) for $L_{\rm {[OIII]}}^{\rm cor}$ (broad) -- $\nu L^{\rm obs}_{\nu}$ (22 $\micron$), $L_{\rm {[OIII]}}^{\rm cor}$ (narrow) -- $\nu L^{\rm obs}_{\nu}$ (22 $\micron$), and $L_{\rm {[OIII]}}^{\rm cor}$ (broad+narrow) -- $\nu L^{\rm obs}_{\nu}$ (22 $\micron$) relations are $\sim$ 0.80, 0.46, and 0.75 with null hypothesis probabilities $P \sim 5.21 \times 10^{-7}$, $2.12 \times 10^{-2}$, and $9.56 \times 10^{-7}$, respectively.
This means that $\nu L^{\rm obs}_{\nu}$ (22 $\micron$) is well-correlated with broad component of \oiii luminosity.   
Note that the SEDs of our IR-bright DOG sample at around 22 $\micron$ appears flat as shown in Figure \ref{SED} \citep[see also][]{Toba_17b}.
Given the somewhat narrow redshift range of our sample (0.05 $< z <$ 1.02), the luminosity in the MIR regime is roughly constant, which would result in a correlation even when using the observed-frame 22 $\micron$ luminosity.
Since 22 $\micron$ luminosity could trace AGN activity and the broad component is likely to be more strongly affected by AGN outflows compared to the narrow component, broad component of \oiii outflow tends to have better correlation with 22 $\micron$ luminosity.
We should keep in mind that, at the same time, the above correlation may be applicable only for IR-bright DOGs because whether or not other population follows this relation is still unknown.

%------------------------
%     VVD diagram
%------------------------    
\subsection{VVD diagram}
\label{sVVD}
Here we show the \oiii velocity offset with respect to the systemic velocity and velocity dispersion diagram (hereafter VVD diagram) for our IR-bright DOG sample and the SDSS Seyfert 2 galaxy sample taken from \cite{Woo}, who investigated outflow properties using a large sample of $\sim$ 40,000 Sy2s at $z< 0.3$.
Figure \ref{VVD} shows the resultant VVD diagram of IR-bright DOGs and SDSS Sy2s where objects only with $\delta_{V_{\rm [OIII]}} < V_{\rm [OIII]}$ and $\delta_{\sigma_{\rm [OIII]}} < \sigma_{\rm [OIII]}$ are plotted.
We found that 16/20 (80\%) DOGs show blueshifted [O{\,\sc iii}], which supports the biconical outflow model combined with dust extinction suggested by \cite{Crenshaw} \citep[see also][]{Barrows}; the redshifted component of outflow (receding cone) tends to be easily hidden by foreground dust.
However, this fraction (0.80) is larger than that of SDSS Sy2s (0.56) with $v_{\rm [OIII]}$ measurements better than 1$\sigma$ probably because receding component of outflowing gas in DOGs is more preferentially hidden by large amount of dust. 
Although the dust geometry between DOGs and Sy2s could be different, it is easy for DOGs to hide the receding outflow than approaching outflows to the line-of-sight.
We also found that the majority of the IR-bright DOGs lie above the SDSS Sy2 on the VVD diagram.
These results could indicate that the IR-bright DOGs are associated with stronger ionized gas outflow (but see Section \ref{VVD_IR}). \\

%%%%%%%%%%%%%%%%%%%%%%
%	  DISCUSSIONS
%%%%%%%%%%%%%%%%%%%%%%
\section{Discussions}
\label{Dis}
%------------------------
% VVD diagram as a function of IR luminosity
%------------------------  
\subsection{VVD diagram as a function of IR luminosity}
\label{VVD_IR}
In Section \ref{sVVD}, we found the IR-bright DOG sample has larger velocity offset and dispersion than those of SDSS Sy2 sample on the VVD diagram.
However, one caution is that more luminous AGN could drive stronger outflow, i.e., we have to compare outflow properties with fixed AGN luminosity.
Since \cite{Toba_16} derived IR luminosity contributed from AGN, $L_{\rm IR}$ (AGN), using the SED fitting for IR-bright DOG sample, we estimated the $L_{\rm IR}$ (AGN) also for the SDSS Sy2 sample.
In order to derive precise IR luminosity contributed from AGN, we compiled far-IR (FIR) data using {\it AKARI} \citep{Murakami} Far-Infrared Surveyor (FIS: \citealt{Kawada}) bright source catalogue (BSC) version 2.0  (I. Yamamura et al. in preparation).
We selected about 400 objects with 65, 90, 140, and 160 $\micron$ data from SDSS Sy2 sample in \cite{Woo}, and conducted the SED fitting with {\tt SEABASs} in the same manner as \cite{Toba_17b}.
Note that we confirmed that the resultant IR luminosity based on this method is consistent with those in local SDSS galaxies selected from \cite{Salim} \citep[see ][in detal]{Toba_17b}. 

   \begin{figure}
   \centering
   \includegraphics[width=0.45\textwidth]{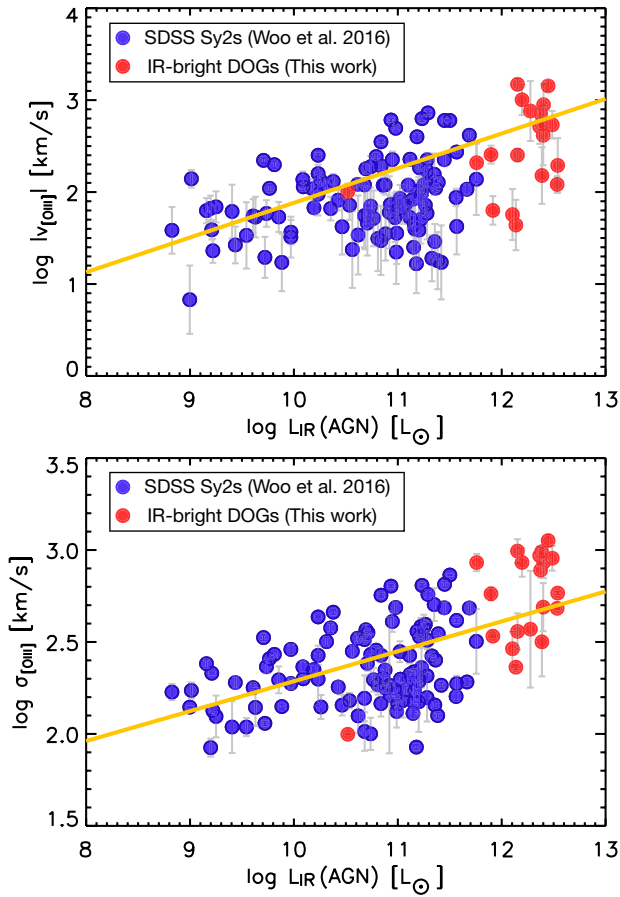}
   \caption{The absolute value of the velocity offset (top) and dispersion (bottom) as a function of IR luminosity contributed from AGN for IR-bright DOG sample (red) and SDSS Sy2 sample (blue). The yellow lines represent the best-fit linear function for both data.}
   \label{vOIII_LAGN}
   \end{figure}
   
Figure \ref{vOIII_LAGN} shows the absolute value of the velocity offset and velocity dispersion as a function of IR luminosity contributed from AGNs for IR-bright DOGs and SDSS Sy2s.
We found that they are continuously distributed on those planes, and $L_{\rm IR}$ (AGN) is well-correlated both with velocity offset and dispersion.
We obtained the following correlation formulae:
\begin{eqnarray}
\log |v_{\rm [OIII]}| & = & (0.377 \pm 0.002) \log L_{\rm IR} \,\, {\rm (AGN)} \nonumber \\
			        & - & (1.887 \pm 0.024),  \\
\log \sigma_{\rm [OIII]} & = & (0.163 \pm 0.001) \log L_{\rm IR} \,\, {\rm (AGN)} \nonumber \\
			         & + &   (0.660 \pm 0.012).  
\end{eqnarray}
Also, our IR-bright DOG sample is basically brighter than Sy2 galaxies, which means that the offset of IR-bright DOG sample compared to SDSS Sy2 sample on the VVD diagram shown in Figure \ref{VVD} is likely due to the difference of IR luminosity originating from AGN activity.

It should be noted that the velocity offset or velocity dispersion itself is not always a good indicator of the strength of AGN outflows because they are affected by dust extinction \citep{Bae_16}.
However, the influence of dust extinction can be minimized if we use the following quantity \citep{Bae_16,Bae_17}; 
\begin{equation}
\sigma_0 = \sqrt{v_{\rm [OIII]}^2 + \sigma_{\rm [OIII]}^2}.
\end{equation}

   \begin{figure*}
   \centering
   \includegraphics[width=0.75\textwidth]{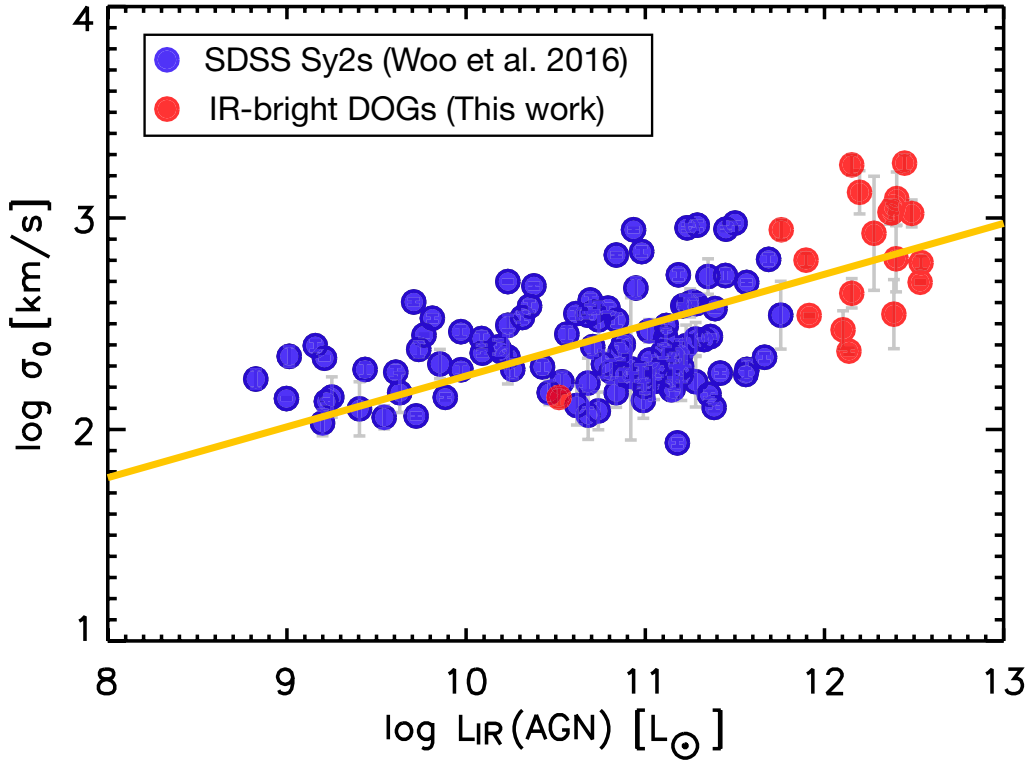}
   \caption{$\sigma_0$ as a function of IR luminosity contributed from AGN for IR-bright DOG sample (red) and SDSS Sy2 sample (blue). The yellow lines represent the best-fit linear function for both data.}
   \label{sigma}
   \end{figure*}

Figure \ref{sigma} shows the relation between $\sigma_0$ and AGN luminosity.
They are well correlated with each other and we obtained the following correlation formula:
\begin{equation}
\log \sigma_0 =  (0.241 \pm 0.001) \log L_{\rm IR} \,\, {\rm (AGN)} 
			        -  (0.152 \pm 0.011).     
\end{equation}
The Spearman rank correlation coefficients ($r_{\rm s}$) for $L_{\rm IR}$ (AGN) -- $v_{\rm [OIII]}$, $L_{\rm IR}$ (AGN) -- $\sigma_{\rm [OIII]}$, and $L_{\rm IR}$ (AGN) -- $\sigma_0$ are $\sim$ 0.51, 0.51, and 0.54 with null hypothesis probabilities $P \sim 7.04 \times 10^{-10}$, $4.88 \times 10^{-10}$, and $6.10 \times 10^{-11}$, respectively. 
We confirmed that the correlation between $\sigma_0$ and $L_{\rm IR}$ (AGN) is slightly stronger than that of others.
Note that \cite{Woo} reported that $\sigma_0$ of SDSS Sy2s correlates with \oiii luminosity where they used \oiii luminosity as an indicator of AGN luminosity.
We conclude that more luminous AGN traced by $L_{\rm IR}$ (AGN) or $L_{\rm [OIII]}$ drives strong outflows. 
\cite{Wagner_11} conducted hydrodynamical simulations of AGN feedback in gas-rich galaxies and concluded that outflow velocities and dispersions of energy driven outflows are determined by the power of the AGN, and all the scatter is determined by the properties of the interstellar medium (ISM) properties, in particular the column density of clumpy gas \citep[see also][]{Wagner_13}.
\cite{Bieri} showed with radiation hydrodynamic simulations of AGN outflows that, for radiation driven winds, the infrared photons provide most of the mechanical advantage to drive outflows to high velocities, and that the properties of the outflows evolved according to the optical depth of infrared photons.
Our observational results support the above conclusions. 

   \begin{figure*}
   \centering
   \includegraphics[width=0.85\textwidth]{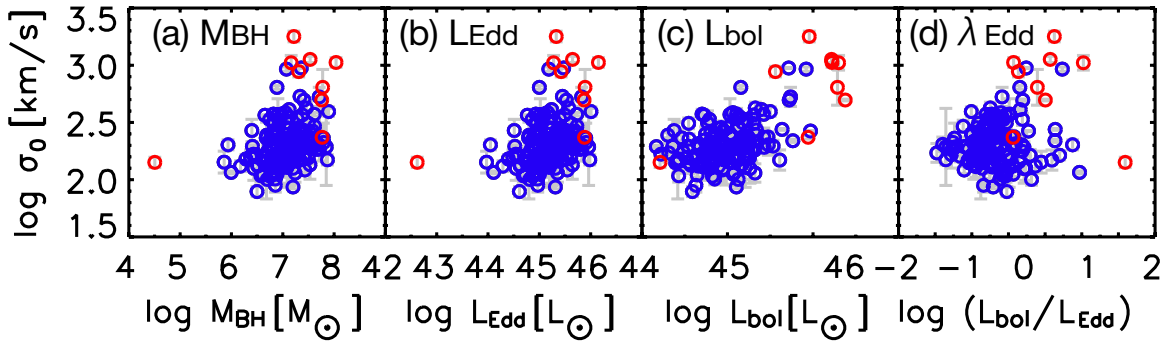}
   \caption{$\sigma_0$ as functions of (a) black hole mass, (b) Eddington luminosity, (c) bolometric luminosity, and (d) Eddington ratio. Symbols are the same as those in Figure \ref{sigma}.}
   \label{sigma_various}
   \end{figure*}

%------------------------
% sigma_0 as a function of other properties
%------------------------  
\subsection{$\sigma_0$ as a function of other properties}
In Section \ref{VVD_IR}, we found that $\sigma_0$, an indicator of the strength of an AGN outflow, depends on $L_{\rm IR}$ (AGN).
Here we investigate the dependence of $\sigma_0$ on other physical quantities; the black hole mass ($M_{\rm BH}$), Eddington luminosity ($L_{\rm Edd}$), bolometric luminosity ($L_{\rm bol}$), and Eddington ratio ($\lambda_{\rm Edd} \equiv L_{\rm bol}/L_{\rm Edd}$).
The black hole mass is estimated from the stellar mass ($M_*$) by using an empirical relation reported in \cite{Reines}; $\log (M_{\rm BH}/M_{\sun})$ = 1.05 $\log \,(M_*/10^{11} M_{\sun})$ + 7.45 with a scatter of 0.24 dex.
The stellar mass is estimated using {\tt SEABASs} in which we employed synthetic stellar templates from \cite{Bruzual} stellar population models assuming a \cite{Chabrier} initial mass function (IMF), and  reddening using a \cite{Calzetti} dust extinction law \citep[see also][]{Toba_17b}.
The Eddington luminosity in units of erg s$^{-1}$ is estimated using $L_{\rm Edd} = 1.3 \times 10^{46} \,(M_{\rm BH}/10^8 M_{\sun})$ \citep{Ferrarese}.
The bolometric luminosity is estimated by integrating the best-fit SED template output by {\tt SEABASs} over wavelengths longward of Ly$\alpha$ in the same manner as \cite{Assef}.
Note that the mean of $L_{\rm bol}$/$L_{\rm IR}$ for IR-bright DOG is 1.61 $\pm$ 0.27, which is consistent with that reported in \cite{Fan}.

Figure \ref{sigma_various} shows $\sigma_0$ as functions of black hole mass, Eddington luminosity, bolometric luminosity, and Eddington ratio.
For any of these quantities, the values of $\sigma_0$ of IR-bright DOGs tend to be larger than those of Sy2.
However, the correlations of these quantities with $\sigma_0$ are not strong compared to the correlation of $L_{\rm IR}$ (AGN) with $\sigma_0$.
Their Spearman rank correlation coefficients are less than 0.4, which could indicate that $M_{\rm BH}$, $L_{\rm Edd}$, $L_{\rm bol}$, and $\lambda_{\rm Edd}$ is unlikely to be a primal parameter while $L_{\rm IR}$ (AGN) is a primal parameter tracing the outflow strength.

%------------------------
% Energetics of AGN outflows
%------------------------
\subsection{Energetics of AGN outflows}
We discuss the energetics of AGN-driven outflows in terms of the mass outflow rate, energy injection rate, and momentum flux of our IR-bright DOG sample.
However, an accurate estimate of these quantities is challenging because such estimates require detailed kinematic modeling for each object.
We thus adopt a simple outflow model for the entire sample to provide first order constraints on
the energetics of IR-bright DOGs.

If we assume a spherical volume of outflowing ionized gas \citep[e.g.,][]{Harrison,Bae_17}, the mass outflow rate ($\dot{M}_{\rm out}$), energy injection rate ($\dot{E}_{\rm out}$), and momentum flux ($\dot{P}_{\rm out}$) are given by 
\begin{eqnarray}
\label{M}
\dot{M}_{\rm out}	&	=	& \frac{3 M_{\rm gas} v_{\rm out}}{R_{\rm out}}	,\\
\dot{E}_{\rm out}	&	=	& \frac{1}{2} \dot{M}_{\rm out} v_{\rm out}^2	,\\
\dot{P}_{\rm out}	&	=	& \dot{M}_{\rm out} v_{\rm out},	
\end{eqnarray}
where $M_{\rm gas}$ is the ionized gas mass, $R_{\rm out}$ is the outflow radius, and $v_{\rm out}$ is the flux-weighted intrinsic outflow velocity or bulk velocity of the outflows.
Assuming case B recombination, the mass of H$\beta$ emitting gas can be estimated as follows \citep{Nesvadba}:
\begin{equation}
M_{\rm gas} = 2.82\times 10^9 \left( \frac{L_{{\rm H}\beta}}{10^{43}\,{\rm erg}\,{\rm s}^{-1}} \right) \left(\frac{n_{\rm e}}{100\, {\rm cm}^{-3}} \right)^{-1},
\end{equation} 
where $L_{\rm H\beta}$ is H$\beta$ luminosity in units of erg s$^{-1}$ and $n_{\rm e}$ is the electron density in unites of cm$^{-3}$.
In this work, we adopt $n_{\rm e}$ = 100 cm$^{-3}$ as routinely assumed in similar works \citep[e.g.,][]{Liu_b,Brusa} and this value is roughly consistent with that derived from \sii doublet in a luminous obscured quasar at $z \sim 1.5$ \citep{Perna}.
For $R_{\rm out}$, we first estimate the size of the narrow line region ($R_{\rm NLR}$) by using an empirical relation between $R_{\rm NLR}$ and extinction--uncorrected \oiii luminosity reported by \cite{Bae_17},
\begin{equation}
\log \,R_{\rm NLR} = (0.41 \pm 0.02) \log \,L_{\rm [OIII]} - (14.00 \pm 0.77).
\end{equation}
We then simply choose $R_{\rm out} = 2 R_{\rm NLR}$ \citep{Bae_17}.
For $v_{\rm out}$, we also use an empirical relation between $v_{\rm out}$ and $\sigma_0$ reported by \cite{Bae_17},
\begin{equation}
\label{v}
v_{\rm out} = (2.0 \pm 0.5) \sigma_0.
\end{equation}
We caution that the electron density depends on the object and $v_{\rm out}$ depends on the dust extinction and inclination of each object \citep[see][and references therein]{Greene,Harrison,Bae_17}, which means that the derived quantities under our simple assumptions could induce large uncertainties.
The resultant values estimated using Equation (\ref{M})--(\ref{v}) are summarized in Table \ref{table2}.

   \begin{figure}
   \centering
   \includegraphics[width=0.45\textwidth]{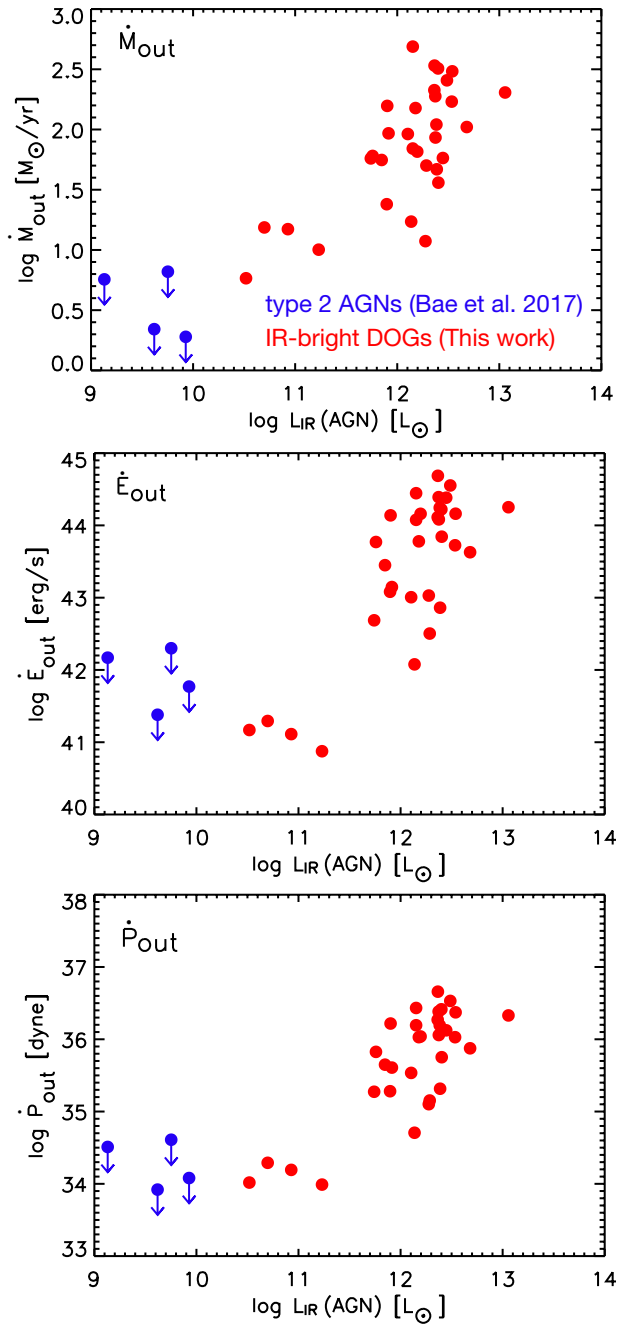}
   \caption{The mass outflow rate $\dot{M}_{\rm out}$ (left), the energy injection rate $\dot{E}_{\rm out}$ (middle), and momentum flux $\dot{P}_{\rm out}$ (right) as a function of IR luminosity contributed from AGNs ($L_{\rm IR}$ (AGN)) of IR-bright DOGs (red circle) and  type 2 AGNs at $z < 0.1$ (blue circle) presented by \cite{Bae_17}.}
   \label{energy}
   \end{figure}

\floattable
\begin{deluxetable}{lcccccccc}
\tabletypesize{\scriptsize}
\tablewidth{0pt} 
\tablecaption{Energetics of WISE-SDSS spec DOGs. \label{table2}}
\tablecolumns{9}
\tablenum{2}
\tablewidth{0pt}
\tablehead{
\colhead{ID}&
%\colhead{objname} &
\colhead{$\log L_{{\rm H}\beta}$} &
\colhead{$\log M_{\rm gas}$} &
\colhead{$\log v_{\rm out}$} &
\colhead{$\log R_{\rm out}$} &
\colhead{$\log \dot{M}_{\rm out}$} &
\colhead{$\log \dot{E}_{\rm out}$} &
\colhead{$\log \dot{P}_{\rm out}$} &
\colhead{$\frac{\dot{P}_{\rm out}}{L_{\rm IR}\,{\rm(AGN)}/c}$} 
\\
\colhead{} &
%\colhead{} &
\colhead{erg s$^{-1}$} &
\colhead{$M_\sun$} &
\colhead{km s$^{-1}$} &
\colhead{pc} &
\colhead{$M_\sun$ yr$^{-1}$} &
\colhead{erg s$^{-1}$} &
\colhead{dyne} &
\colhead{} 
}
\startdata
 1 & 41.7 &  8.2 &  2.3 &  3.8 &  1.2 & 41.3 & 34.3 &  3.0\\
 2 & 41.3 &  7.8 &  3.6 &  4.1 &  1.8 & 44.4 & 36.1 &  3.7\\
 3 & 41.4 &  7.9 &  2.8 &  3.5 &  1.7 & 42.9 & 35.3 &  0.7\\
 4 & 41.2 &  7.7 &  2.7 &  3.6 &  1.2 & 42.1 & 34.7 &  0.3\\
 5 & 41.7 &  8.1 &  3.4 &  3.9 &  2.0 & 44.2 & 36.2 &  5.0\\
 6 & 41.3 &  7.8 &  3.3 &  3.7 &  1.9 & 44.1 & 36.1 &  3.8\\
 7 & 41.7 &  8.2 &  3.2 &  3.7 &  2.2 & 44.1 & 36.2 & 16.1\\
 8 & 42.9 &  9.4 &  2.9 &  4.1 &  2.7 & 44.1 & 36.4 & 14.9\\
 9 & 42.2 &  8.7 &  2.8 &  3.9 &  2.0 & 43.0 & 35.5 &  2.1\\
10 & 42.0 &  8.5 &  3.1 &  4.0 &  2.0 & 43.6 & 35.9 &  1.2\\
11 & 41.5 &  7.9 &  2.2 &  3.6 &  1.0 & 40.9 & 34.0 &  0.4\\
14 & 41.1 &  7.6 &  3.6 &  3.7 &  1.8 & 44.4 & 36.2 &  8.6\\
15 & 42.4 &  8.8 &  3.1 &  3.9 &  2.5 & 44.2 & 36.4 &  8.1\\
16 & 40.5 &  6.9 &  3.2 &  3.6 &  1.1 & 43.0 & 35.1 &  0.5\\
17 & 40.6 &  7.1 &  3.1 &  3.3 &  1.4 & 43.1 & 35.3 &  1.9\\
18 & 42.6 &  9.1 &  3.1 &  4.2 &  2.5 & 44.2 & 36.4 &  5.3\\
19 & 42.3 &  8.8 &  3.1 &  4.1 &  2.3 & 44.1 & 36.3 &  6.3\\
20 & 42.4 &  8.9 &  3.2 &  4.3 &  2.3 & 44.3 & 36.3 &  1.5\\
21 & 41.6 &  8.0 &  3.1 &  3.9 &  1.7 & 43.4 & 35.6 &  4.9\\
23 & 41.8 &  8.2 &  2.2 &  3.7 &  1.2 & 41.1 & 34.2 &  1.4\\
24 & 42.1 &  8.5 &  3.3 &  4.0 &  2.3 & 44.4 & 36.4 &  8.0\\
25 & 41.6 &  8.0 &  2.7 &  3.5 &  1.7 & 42.5 & 35.2 &  0.6\\
27 & 42.2 &  8.6 &  2.8 &  4.0 &  2.0 & 43.1 & 35.6 &  3.8\\
28 & 41.0 &  7.5 &  3.2 &  3.4 &  1.8 & 43.8 & 35.8 &  9.1\\
29 & 42.4 &  8.9 &  3.0 &  4.1 &  2.2 & 43.7 & 36.0 &  2.4\\
30 & 42.2 &  8.7 &  3.3 &  4.1 &  2.4 & 44.6 & 36.5 &  8.6\\
31 & 42.4 &  8.8 &  3.1 &  4.2 &  2.2 & 43.8 & 36.0 &  5.5\\
32 & 41.3 &  7.7 &  3.4 &  3.8 &  1.8 & 44.2 & 36.0 &  5.4\\
33 & 42.0 &  8.5 &  3.3 &  3.7 &  2.5 & 44.7 & 36.7 & 15.3\\
34 & 42.0 &  8.4 &  2.7 &  3.9 &  1.8 & 42.7 & 35.3 &  2.7\\
35 & 40.8 &  7.3 &  3.4 &  3.6 &  1.6 & 43.8 & 35.8 &  1.7\\
36 & 40.6 &  7.0 &  2.5 &  3.2 &  0.8 & 41.2 & 34.0 &  2.5\\
\enddata
\end{deluxetable}

Figure \ref{energy} shows the energetics ($\dot{M}_{\rm out}$, $\dot{E}_{\rm out}$, and $\dot{P}_{\rm out}$) as a function of $L_{\rm IR}$ (AGN) for IR-bright DOGs and type 2 AGNs reported by \cite{Bae_17}. 
\cite{Bae_17} observed type 2 AGNs at $z < 0.1$ with integral-field spectroscopy and investigated the energetics of them.
We estimate their $L_{\rm IR}$ (AGN) based on the SED fitting in the same manner as those we described in Section \ref{VVD_IR} and 4 AGNs are plotted in Figure \ref{energy}. 
We found that our IR-bright DOG sample have systematically larger values than those of local type 2 AGNs.
Since these values are clearly connected to AGN activity as shown in Figure \ref{energy} \citep[see also][]{Bae_17}, this result can be explained by the difference of AGN luminosity as discussed in Section \ref{VVD_IR}.

We also estimate the ``momentum boost'', i.e., the ratio of the momentum flux ($\dot{P}$) and the AGN radiative momentum output ($L_{\rm IR}$ (AGN)/$c$) (see Table \ref{table2}).
We found that the estimated initial velocity ($v_{\rm in}$) from nucleus for most objects assuming that the observed  outflows are energy-conserving \citep[see][]{Faucher-Giguere,Cicone} is $v_{\rm in} = (0.01-0.2)c$.
This result suggests that some IR-bright DOGs show an ultrafast outflow (UFO) with $v_{\rm in} = (0.05-0.3)c$ \citep[e.g.,][]{Tombesi,Gofford}.

%------------------------
% VVD diagram for other lines
%------------------------
\subsection{VVD diagram for other lines}
\begin{figure}
   \centering
   \includegraphics[width=0.45\textwidth]{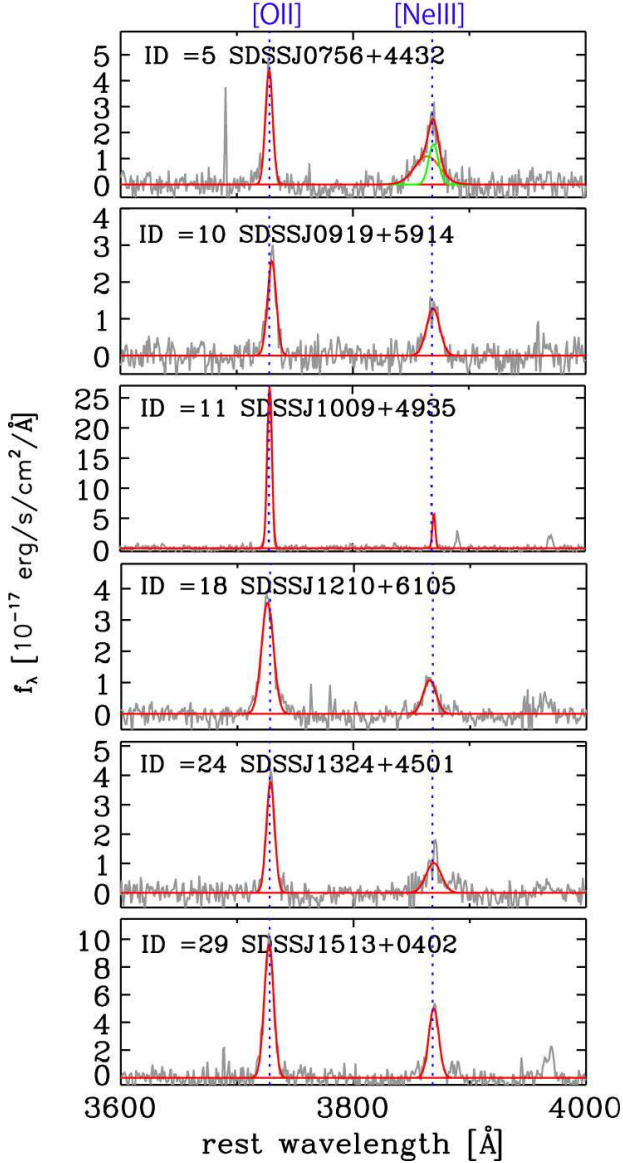}
   \caption{The example of spectral fitting for \oii and \neiii lines. Gray line shows the starlight-subtracted spectra. The red line shows the best fitting with single or double Gaussian for each emission line. The green and orange lines show narrow and broad emission line component, respectively. The vertical blue dashed liens correspond to the rest-frame wavelength for [O{\,\sc ii}] and [Ne{\,\sc iii}] lines.}
   \label{spec_OII_NeIII}
   \end{figure}

We discuss the outflow properties of other emission lines.
Figure \ref{spec_OII_NeIII} shows examples of the spectra fitting for [O{\,\sc ii}]$\lambda3727$ and [Ne{\,\sc iii}]$\lambda3869$ lines.
Both lines are well-fitted by single or double Gaussians.

\begin{figure}
   \centering
   \includegraphics[width=0.45\textwidth]{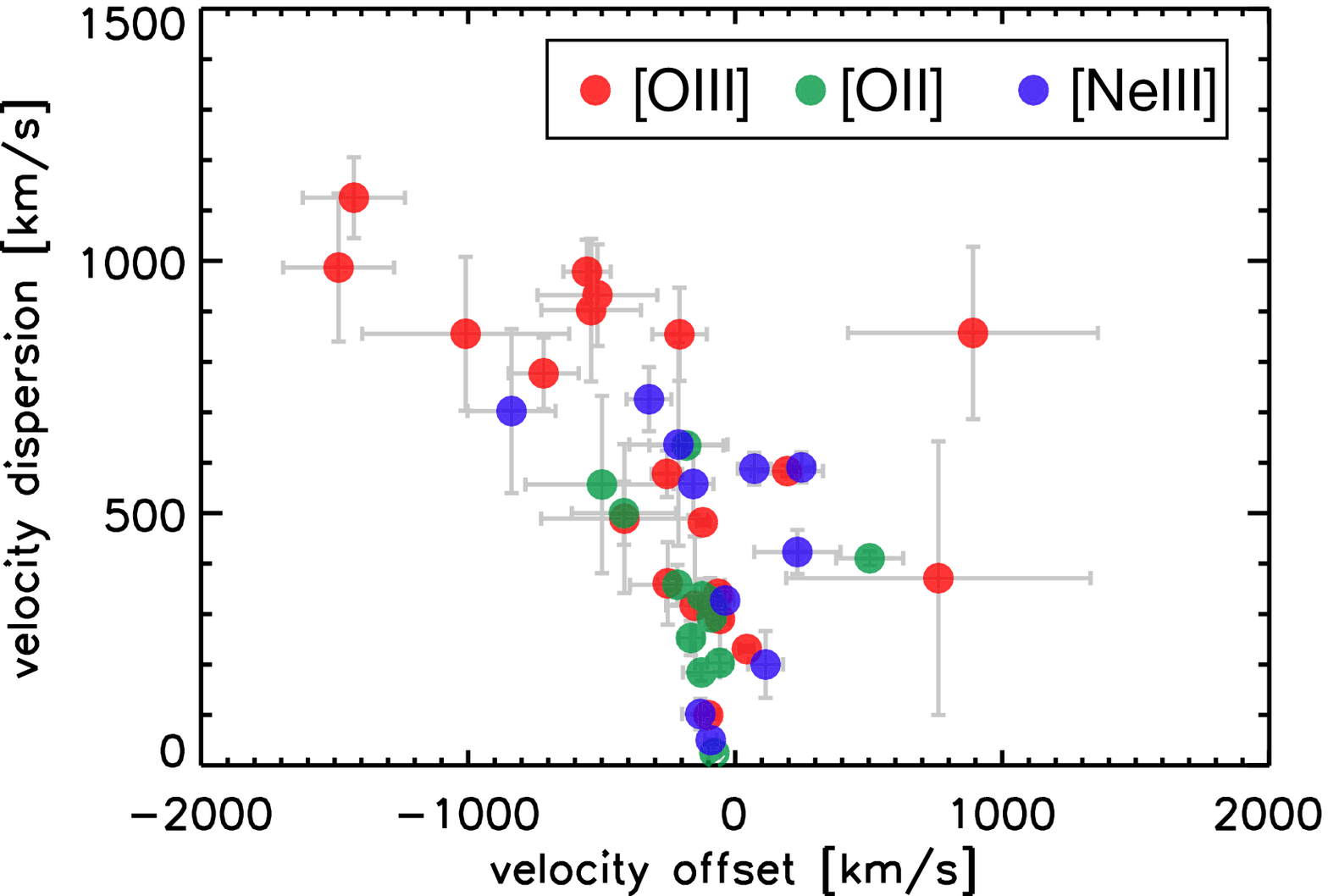}
   \caption{VVD diagram for \oii (green), \neiii (blue), \oiii (red) line.}
   \label{VVD_OII_NeIII}
   \end{figure}

Figure \ref{VVD_OII_NeIII} show the VVD diagram for [Ne{\,\sc iii}], [O{\,\sc ii}], and \oiii for our IR-bright DOG sample.
We found that \neiii have similar velocity offset and dispersion as those of \oiii while \oii have smaller values than those of [O{\,\sc iii}]. 
It should be noted that \oii is not well fitted with double Gaussian component in many cases due to the blending of $\lambda$3726, 3729 \AA \,\,doublet.
If we use only a single Gaussian, alternatively, it gives a lot larger velocity dispersion  ($\sigma_{\rm [OII]}$).
We should keep in mind the above uncertainties before interpreting the discrepancy between \neiii and [O{\,\sc iii}], and \oii in Figure \ref{VVD_OII_NeIII}.

The difference of $v_{\rm line}$ and $\sigma_{\rm line}$ for each line tells us a hint to understand the physicochemical properties of outflowing gas.
The ionization potentials of [O{\,\sc ii}]$\lambda$3727, [O{\,\sc iii}]$\lambda$5007, and [Ne{\,\sc iii}]$\lambda$3869 are 13.61, 35.15, and 41.07 eV, respectively.
The critical electron densities for collisional de-excitation of [O{\,\sc ii}]$\lambda$3727, [O{\,\sc iii}]$\lambda$5007, and [Ne{\,\sc iii}]$\lambda$3869 are $3.4 \times 10^3$, $6.8 \times 10^5$, and $9.5 \times 10^6$ cm$^{-3}$, respectively.  
The fraction of objects with $|v_{\rm line}| > 50$ km s$^{-1}$ and $\sigma_{\rm line} > 500$ km s$^{-1}$ for [O{\,\sc ii}], [O{\,\sc iii}], and [Ne{\,\sc iii}] are 0.134, 0.566, and 0.571, respectively.
This means that more dense and ionized gas tend to show larger velocity offset and dispersion. 
Since it is naturally expected that electron densities will increase toward the nuclear region and gas located there is highly ionized by AGN radiation, \oiii and \neiii are ejected with high velocity while \oii are less affected by AGN radiation, that picture is consistent with those suggested by \cite{Barrows} \citep[see also][]{Komossa}.

   \begin{figure}
   \centering
   \includegraphics[width=0.45\textwidth]{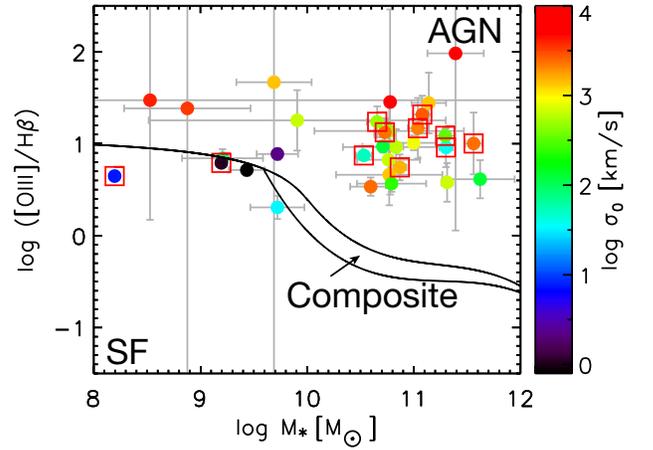}
   \caption{MEx diagnostic diagram for the IR-bright DOGs. The color scheme indicates $\sigma_0 = \sqrt{v_{\rm [OIII]}^2 + \sigma_{\rm [OIII]}^2}$. The two solid lines curves are empirically determined to distinguish AGN, SF, and composite type of galaxies. The data with red square have SN $>$ 3 both for [O{\,\sc iii}]/H$\beta$ and stellar mass.}
   \label{MEx}
   \end{figure}   
   
%------------------------
%    MEx diagram
%------------------------
\subsection{MEx diagram}
Finally, we discuss the Mass-Excitation (MEx) diagram \citep{Juneau_11,Juneau_14} that enables to perform AGN diagnostics for objects with even $z > 0.4$.
Since {\tt SEABASs} outputs stellar mass ($M_*$) and we measured \oiii and H$\beta$ line flux, we here investigate where IR-bright DOGs lie in the MEx diagram.
Note the our estimate based on this method have an uncertainty because we did not take into account the influence from the scattered light by AGNs \citep[see also][]{Hamann,Toba_17b}.
We also note that we excluded DOGs classified as type 1 AGN (see Section \ref{DA}) in this analysis because MEx diagram is optimized for galaxies/AGNs with narrow line emission.

Figure \ref{MEx} shows the MEx diagnostic diagram for the IR-bright DOGs, suggesting that IR-bright DOGs can be classified as AGNs, which is consistent from our inspection based on the SED and IR flux dependence of the AGN fraction for DOGs \citep[see ][]{Toba_15,Toba_16}.
At the same time, there are no significant dependences of $\sigma_0$ on the MEx diagram.
This could indicate that [OIII]/H$\beta$ is unlikely to a good tracer of outflow strength partly because [OIII]/H$\beta$ also depends on other quantities such as metallicity.
On the other hand, after removing data with large error, i.e., when focusing only on data with SN $>$ 3 both for [O{\,\sc iii}]/H$\beta$ and stellar mass, stellar mass is likely to be correlated with $\sigma_0$.
Since stellar mass correlates with stellar dispersion that also correlates with $\sigma_0$ \citep{Woo}, this tendency is naturally expected.\\

%%%%%%%%%%%%%%%%%%%%%%
%	  SUMMARY
%%%%%%%%%%%%%%%%%%%%%%
\section{Summary}
\label{Sum}
In this work, we investigated the outflowing ionized gas properties of IR-bright DOGs by performing  detailed spectral analysis for their SDSS spectra.
Among 67 IR-bright DOGs selected with the {\it WISE} and SDSS spectroscopic catalogs, 36 objects show [O{\,\sc iii}]$\lambda$5007 line and we estimated its velocity offset with respect to the systemic velocity and velocity dispersion.
In particular, we conducted spectral fitting with single or double Gaussian component depending on whether or not they have broad wing.
The main results are as follows:
\begin{enumerate}
\item Among a sample of 32 IR-bright DOGs that are classified as type 2 AGN, 24 ($\sim$75\%) objects have large \oiii velocity dispersion with 300 km s$^{-1}$. This fraction is larger than other AGN populations, indicating that IR-bright DOGs show stronger ionized gas outflow.
\item The \oiii luminosity is correlated with observed-frame luminosity at 22 $\micron$. In particular, the 22 $\micron$ luminosity at observed-frame may be a good indicator of the luminosity of broad component of \oiii line for IR-bright DOGs.
\item The infrared luminosity contributed from AGNs for IR-bright DOG + SDSS Seyfert 2 sample is well-correlated with velocity offset ($v_{\rm [OIII]}$), dispersion ($\sigma_{\rm [OIII]}$), and particularly $\sigma_0 = \sqrt{v_{\rm [OIII]}^2 + \sigma_{\rm [OIII]}^2}$.
This indicates that objects with higher AGN luminosity tend to launch stronger outflowing gas.
\item IR-bright DOG sample have larger velocity offset and dispersion than those of the SDSS Seyfert 2 sample, which can be interpreted as the difference of their AGN luminosities.
\item The energetics ($\dot{M}_{\rm out}$, $\dot{E}_{\rm out}$, and $\dot{P}_{\rm out}$) of IR-bright DOGs correlates with AGN luminosity. 
Some IR-bright DOGs have initial outflow velocity ($v_{\rm in}$) $>$ 0.1$c$, which means that some IR-bright DOGs show an ultrafast outflow.
\item The velocity offset and dispersion of \oiii and [Ne{\,\sc iii}]$\lambda3869$ are larger than those of [O{\,\sc ii}]$\lambda3727$, suggesting that denser and more ionized gas are effectively affected by AGN radiation.
\end{enumerate}

%%%%%%%%%%%%%%%%%%%%%%%%%
%%%%%%%%%%%%%%%%%%%%%%%%%
\acknowledgments
The authors gratefully acknowledge the anonymous referee for a careful reading of the manuscript and very helpful comments.
Funding for SDSS-III has been provided by the Alfred P. Sloan Foundation, the Participating Institutions, the National Science Foundation, and the U.S. Department of Energy Office of Science. The SDSS-III web site is http://www.sdss3.org/.
SDSS-III is managed by the Astrophysical Research Consortium for the Participating Institutions of the SDSS-III Collaboration including the University of Arizona, the Brazilian Participation Group, Brookhaven National Laboratory, Carnegie Mellon University, University of Florida, the French Participation Group, the German Participation Group, Harvard University, the Instituto de Astrofisica de Canarias, the Michigan State/Notre Dame/JINA Participation Group, Johns Hopkins University, Lawrence Berkeley National Laboratory, Max Planck Institute for Astrophysics, Max Planck Institute for Extraterrestrial Physics, New Mexico State University, New York University, Ohio State University, Pennsylvania State University, University of Portsmouth, Princeton University, the Spanish Participation Group, University of Tokyo, University of Utah, Vanderbilt University, University of Virginia, University of Washington, and Yale University.
This publication makes use of data products from the Wide-field Infrared Survey Explorer, which is a joint project of the University of California, Los Angeles, and the Jet Propulsion Laboratory/California Institute of Technology, funded by the National Aeronautics and Space Administration. 
This research is based on observations with AKARI, a JAXA project with the participation of ESA.
This research has made use of the NASA/ IPAC Infrared Science Archive, which is operated by the Jet Propulsion Laboratory, California Institute of Technology, under contract with the National Aeronautics and Space Administration.
Y.Toba and W.H.Wang acknowledge the support from the Ministry of Science and Technology
of Taiwan (MOST 105-2112-M-001-029-MY3).
T.Nagao is financially supported by the Japan Society for the Promotion of Science (JSPS) KAKENHI (16H01101 and 16H03958).
J.H.Woo acknowledges the support by the National Research Foundation of Korea grant funded by the Korea government (No. 2016R1A2B3011457).

\appendix
\section{The SDSS spectra of IR-bright DOGs with a powerful \oiii outflow}
\label{spectra_all}
\begin{figure}
   \centering
   \includegraphics[width=0.85\textwidth]{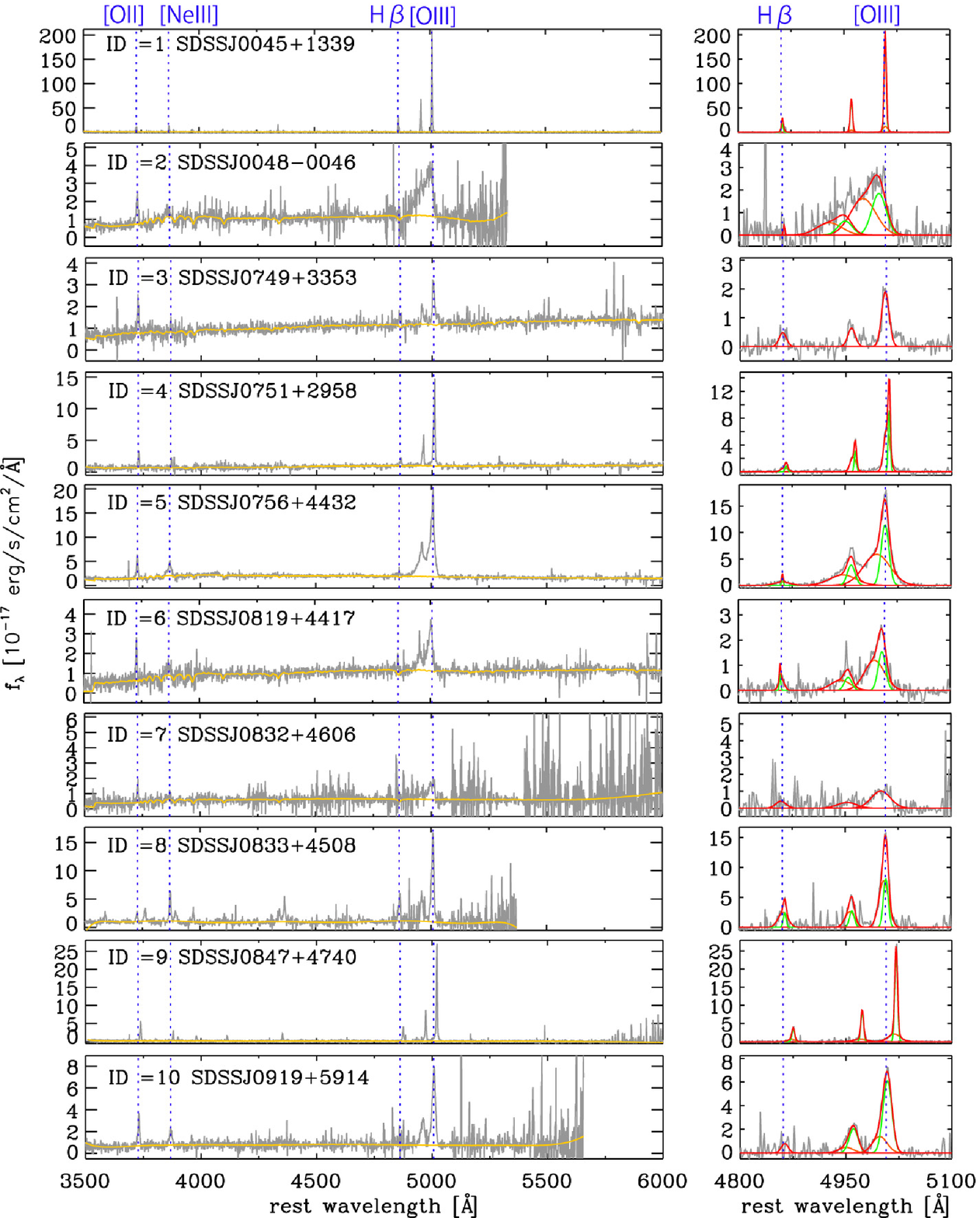}
   \caption{The SDSS spectra for IR-bright DOG sample with ID=1--10. The yellow lines show our best fits to the continuum. The right panel for each object shows the starlight-subtracted spectra with gray solid line. The red line shows the best fitting with single or double Gaussian for each emission line. The green and orange lines show narrow and broad emission line component for each double Gaussian, respectively. The vertical blue dashed liens correspond to the rest-frame wavelength for [O{\,\sc ii}], [Ne{\,\sc iii}], H$\beta$, and [O{\,\sc iii}] lines.}
   \label{spectra}
   \end{figure}

\begin{figure}
   \centering
   \includegraphics[width=0.85\textwidth]{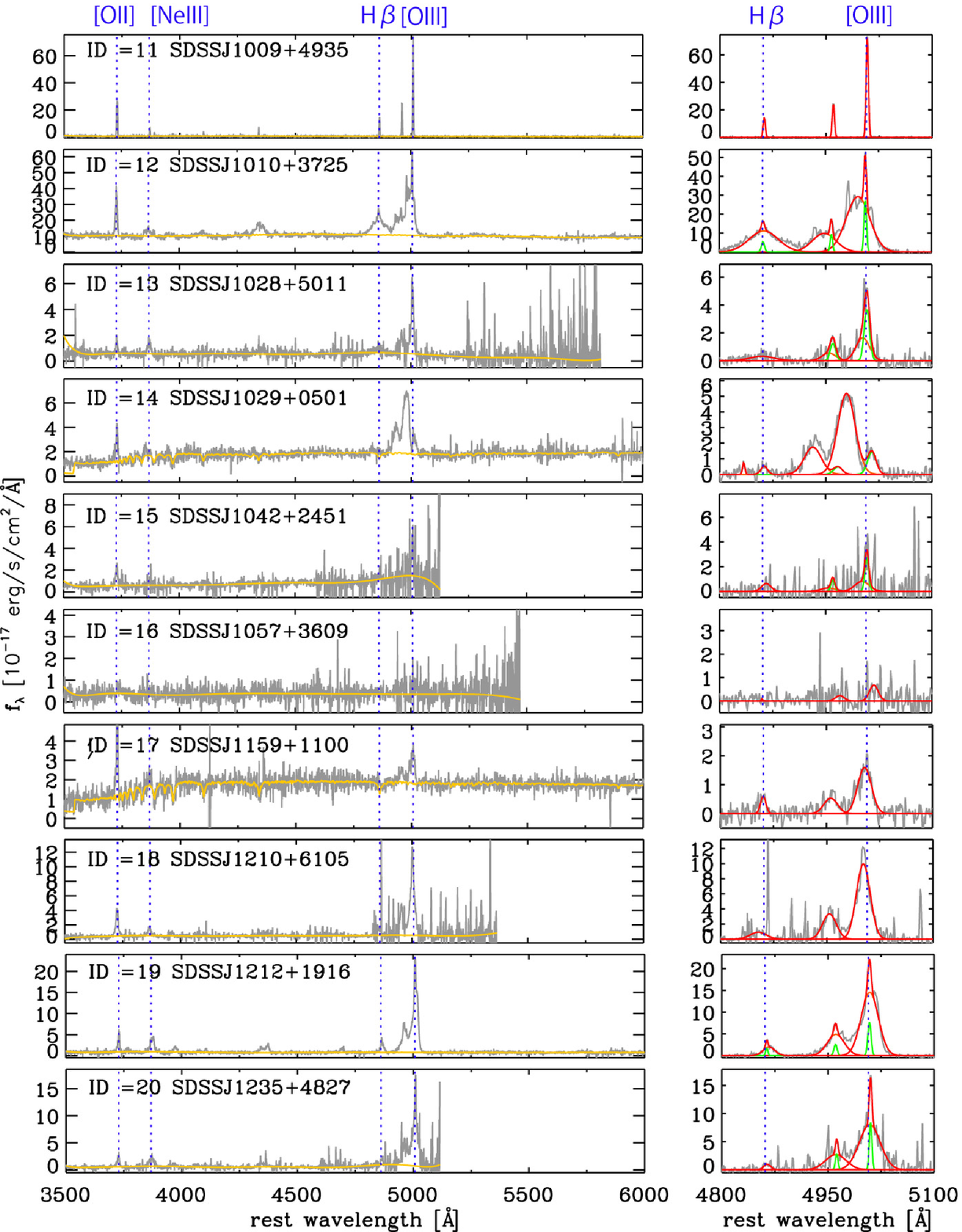}
   \caption{Same as Figure \ref{spectra}, but for DOGs with ID=11--20.}
   \label{spectra2}
   \end{figure}

\begin{figure}
   \centering
   \includegraphics[width=0.85\textwidth]{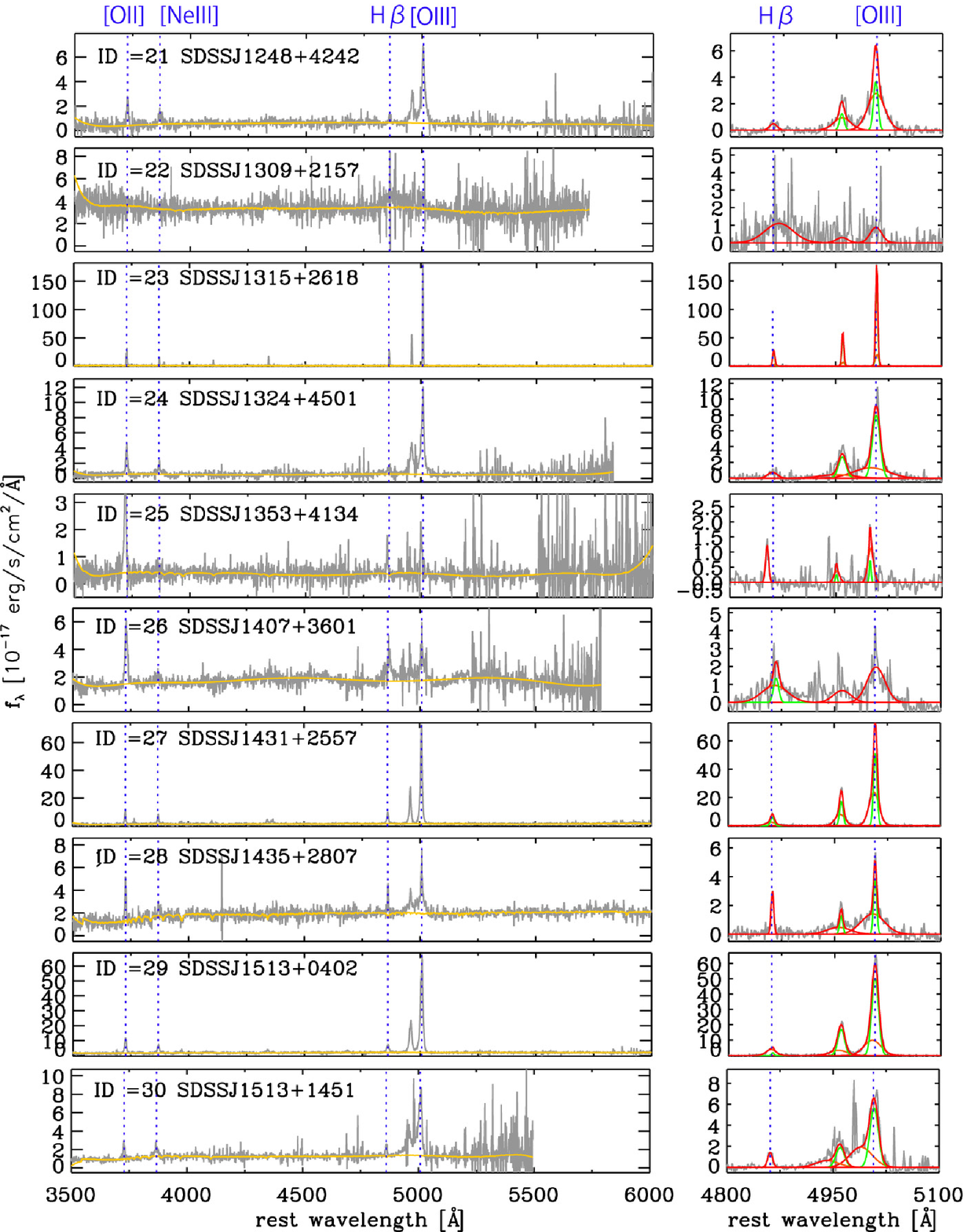}
   \caption{Same as Figure \ref{spectra}, but for DOGs with ID=21--30.}
   \label{spectra3}
   \end{figure}

\begin{figure}
   \centering
   \includegraphics[width=0.85\textwidth]{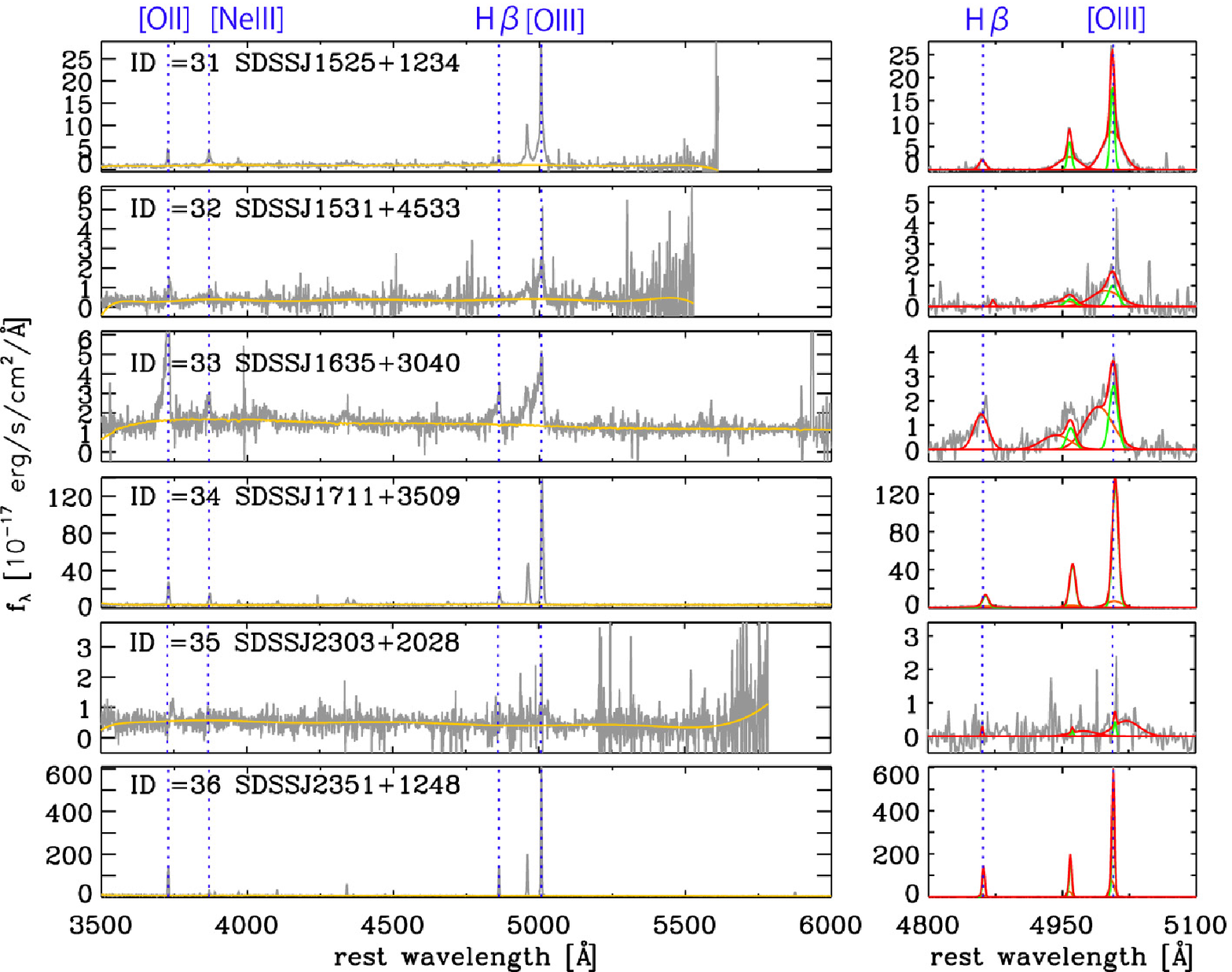}
   \caption{Same as Figure \ref{spectra}, but for DOGs with ID=31--36.}
   \label{spectra4}
   \end{figure}

Here we present the SDSS spectra for all 36 IR-bright DOGs (Figure \ref{spectra}--\ref{spectra4}).
 
 \floattable
\begin{deluxetable*}{ccccrrrrcrrrcrr}
\tabletypesize{\scriptsize}
\rotate
\tablecaption{[OIII] properties of WISE-SDSS spec DOGs. \label{table}}
\tablecolumns{15}
\tablenum{1}
\tablewidth{0pt}
\tablehead{
\colhead{ID}&
\colhead{objname} &
\colhead{R.A.\tablenotemark{a}} &
\colhead{Decl.\tablenotemark{a}} & 
\colhead{Plate} &
\colhead{fiberID} &
\colhead{MJD} &
\colhead{redshift} & 
\colhead{type \tablenotemark{b}} 	   &
\colhead{$i - [22]$} &
\colhead{$\log L_{\rm IR}$\tablenotemark{c}} &
\colhead{$\log L_{\rm bol}$\tablenotemark{d}} &
\colhead{$w_{\rm [OIII]}$ \tablenotemark{e}}	&
\colhead{$v_{\rm [OIII]}$} &
\colhead{$\sigma_{\rm [OIII]}$}
\\
\colhead{} &
\colhead{} &
\colhead{hms} &
\colhead{dms} &
\colhead{} &
\colhead{} &
\colhead{} &
\colhead{} &
\colhead{} &
\colhead{AB mag} &
\colhead{$L_{\sun}$} &
\colhead{erg s$^{-1}$} &
\colhead{} &
\colhead{km/s} &
\colhead{km/s} 
}
\startdata
 1 &    SDSSJ0045+1339 & 00:45:29.1 & +13:39:08.6 &  419 & 137 & 51879 & 0.295 & type 2 & 7.11 & 10.70 & 44.58 & 1 &    16.8  $\pm$   20.8 &   99.3  $\pm$    3.5\\
 2 &    SDSSJ0048-0046 & 00:48:46.4 & -00:46:11.9 & 3590 & 256 & 55201 & 0.939 & type 2 & 7.55 & 12.45 & 46.26 & 1 & -1427.8  $\pm$  191.5 & 1125.4  $\pm$   80.2\\
 3 &    SDSSJ0749+3353 & 07:49:34.6 & +33:53:08.6 & 3751 & 813 & 55234 & 0.620 & type 2 & 7.14 & 12.39 & 46.07 & 1 &  -151.3  $\pm$  107.3 &  316.7  $\pm$  136.9\\
 4 &    SDSSJ0751+2958 & 07:51:20.5 & +29:58:47.1 & 3752 & 435 & 55236 & 0.437 & type 2 & 7.14 & 12.14 & 45.94 & 1 &    43.8  $\pm$   27.6 &  230.6  $\pm$    6.7\\
 5 &    SDSSJ0756+4432 & 07:56:09.9 & +44:32:22.8 & 6376 & 806 & 56269 & 0.510 & type 2 & 7.18 & 12.39 & 46.22 & 1 &  -555.0  $\pm$   88.9 &  978.6  $\pm$   63.4\\
 6 &    SDSSJ0819+4417 & 08:19:47.3 & +44:17:22.8 & 6379 & 933 & 56340 & 0.578 & type 2 & 7.28 & 12.38 & 46.22 & 1 &  -717.2  $\pm$  130.8 &  777.2  $\pm$   70.9\\
 7 &    SDSSJ0832+4606 & 08:32:48.2 & +46:06:02.6 & 5160 & 330 & 55895 & 0.721 & type 2 & 7.10 & 11.90 & 45.73 & 0 &  -229.3  $\pm$  500.1 &  801.5  $\pm$   45.0\\
 8 &    SDSSJ0833+4508 & 08:33:38.5 & +45:08:33.5 & 7326 & 452 & 56710 & 0.925 & type 2 & 7.05 & 12.15 & 46.04 & 1 &  -252.6  $\pm$   38.0 &  360.9  $\pm$   81.9\\
 9 &    SDSSJ0847+4740 & 08:47:15.0 & +47:40:14.0 & 7320 & 160 & 56722 & 0.713 & type 2 & 7.39 & 12.10 & 45.88 & 1 &   -56.9  $\pm$   36.3 &  290.5  $\pm$   62.2\\
10 &    SDSSJ0919+5914 & 09:19:45.0 & +59:14:30.9 & 5712 & 229 & 56602 & 0.829 & type 2 & 7.22 & 12.68 & 46.34 & 1 &  -184.0  $\pm$  199.2 &  536.3  $\pm$   43.0\\
11 &    SDSSJ1009+4935 & 10:09:41.3 & +49:35:26.5 & 7381 & 548 & 56717 & 0.308 & type 2 & 7.19 & 11.23 & 44.90 & 0 &     5.6  $\pm$   24.2 &   76.7  $\pm$    0.6\\
12 &    SDSSJ1010+3725 & 10:10:34.2 & +37:25:14.7 & 1426 & 110 & 52993 & 0.282 & type 1 & 7.23 & 12.07 & 45.89 & 1 &  -613.6  $\pm$   29.5 &  971.1  $\pm$    6.7\\
13 &    SDSSJ1028+5011 & 10:28:01.5 & +50:11:02.5 & 6694 & 430 & 56386 & 0.776 & type 1 & 7.20 & 11.97 & 45.84 & 1 &   -11.9  $\pm$  259.3 &  476.0  $\pm$   45.4\\
14 &    SDSSJ1029+0501 & 10:29:05.9 & +05:01:32.4 & 4772 & 617 & 55654 & 0.493 & type 2 & 7.11 & 12.16 & 45.96 & 0 & -1485.0  $\pm$  207.7 &  986.9  $\pm$  146.6\\
15 &    SDSSJ1042+2451 & 10:42:41.1 & +24:51:07.0 & 6417 & 509 & 56308 & 1.026 & type 2 & 7.02 & 12.40 & 46.29 & 1 &  -414.8  $\pm$  311.8 &  489.2  $\pm$  147.6\\
16 &    SDSSJ1057+3609 & 10:57:14.5 & +36:09:03.3 & 4626 & 442 & 55647 & 0.885 & type 2 & 7.06 & 12.28 & 46.06 & 0 &   761.0  $\pm$  570.1 &  371.1  $\pm$  271.4\\
17 &    SDSSJ1159+1100 & 11:59:15.3 & +11:00:42.8 & 5388 & 398 & 55983 & 0.351 & type 2 & 7.01 & 11.90 & 45.57 & 0 &  -255.5  $\pm$   57.4 &  578.0  $\pm$   46.0\\
18 &    SDSSJ1210+6105 & 12:10:56.9 & +61:05:51.5 & 6972 & 272 & 56426 & 0.926 & type 2 & 7.63 & 12.54 & 46.34 & 1 &   195.1  $\pm$  133.3 &  583.4  $\pm$   10.2\\
19 &    SDSSJ1212+1916 & 12:12:36.5 & +19:16:23.7 & 5848 & 737 & 56029 & 0.620 & type 2 & 7.37 & 12.37 & 46.19 & 1 &    -1.0  $\pm$   48.8 &  695.9  $\pm$   19.5\\
20 &    SDSSJ1235+4827 & 12:35:44.9 & +48:27:15.4 & 6670 & 254 & 56389 & 1.023 & type 2 & 7.50 & 13.06 & 46.70 & 1 &   -26.1  $\pm$  150.3 &  834.7  $\pm$   37.6\\
21 &    SDSSJ1248+4242 & 12:48:36.1 & +42:42:59.3 & 4703 & 632 & 55617 & 0.682 & type 2 & 7.11 & 11.85 & 45.66 & 1 &   -12.8  $\pm$  130.5 &  632.1  $\pm$   45.4\\
22 &    SDSSJ1309+2157 & 13:09:56.3 & +21:57:00.8 & 2650 &  23 & 54505 & 0.609 & type 1 & 7.49 & 12.06 & 45.88 & 0 &  -518.0  $\pm$  348.2 &  485.5  $\pm$  104.2\\
23 &    SDSSJ1315+2618 & 13:15:14.0 & +26:18:41.3 & 2243 & 171 & 53794 & 0.305 & type 2 & 8.03 & 10.93 & 44.77 & 1 &    12.4  $\pm$   20.9 &   82.1  $\pm$    2.6\\
24 &    SDSSJ1324+4501 & 13:24:40.1 & +45:01:33.8 & 6625 & 124 & 56386 & 0.774 & type 2 & 7.78 & 12.38 & 46.20 & 1 &  -118.1  $\pm$  234.9 & 1009.2  $\pm$  103.6\\
25 &    SDSSJ1353+4134 & 13:53:34.6 & +41:34:39.0 & 6631 & 212 & 56364 & 0.686 & type 2 & 7.43 & 12.29 & 45.93 & 1 &    67.2  $\pm$   93.3 &  214.1  $\pm$   44.9\\
26 &    SDSSJ1407+3601 & 14:07:44.0 & +36:01:09.5 & 3854 &  24 & 55247 & 0.783 & type 1 & 7.17 & 12.58 & 46.36 & 0 &  -257.5  $\pm$   56.4 &  754.9  $\pm$   43.4\\
27 &    SDSSJ1431+2557 & 14:31:36.4 & +25:57:06.8 & 2135 & 482 & 53827 & 0.481 & type 2 & 7.26 & 11.92 & 45.72 & 1 &   -63.2  $\pm$   22.5 &  340.1  $\pm$    4.3\\
28 &    SDSSJ1435+2807 & 14:35:40.3 & +28:07:25.5 & 6018 & 975 & 56067 & 0.346 & type 2 & 7.28 & 11.76 & 45.56 & 1 &  -208.7  $\pm$  102.0 &  854.6  $\pm$   92.2\\
29 &    SDSSJ1513+0402 & 15:13:33.8 & +04:02:22.8 & 4776 &  25 & 55652 & 0.597 & type 2 & 7.23 & 12.54 & 46.38 & 1 &  -121.1  $\pm$   24.7 &  481.8  $\pm$    6.2\\
30 &    SDSSJ1513+1451 & 15:13:54.4 & +14:51:25.2 & 5486 & 200 & 56030 & 0.882 & type 2 & 7.33 & 12.49 & 46.30 & 1 &  -539.3  $\pm$  186.3 &  902.4  $\pm$  141.3\\
31 &    SDSSJ1525+1234 & 15:25:04.7 & +12:34:01.7 & 5492 & 818 & 56010 & 0.851 & type 2 & 7.19 & 12.18 & 46.01 & 1 &    10.5  $\pm$   45.9 &  562.5  $\pm$   19.9\\
32 &    SDSSJ1531+4533 & 15:31:05.1 & +45:33:03.4 & 6735 & 246 & 56397 & 0.871 & type 2 & 7.47 & 12.20 & 46.05 & 1 & -1009.2  $\pm$  387.8 &  855.6  $\pm$  152.5\\
33 &    SDSSJ1635+3040 & 16:35:59.3 & +30:40:32.8 & 5202 & 322 & 55824 & 0.578 & type 2 & 7.09 & 12.37 & 46.02 & 1 &  -515.7  $\pm$  224.2 &  932.1  $\pm$  100.7\\
34 &    SDSSJ1711+3509 & 17:11:45.7 & +35:09:27.7 & 4994 & 525 & 55739 & 0.316 & type 2 & 7.33 & 11.74 & 45.58 & 1 &    -5.2  $\pm$   22.0 &  258.8  $\pm$    5.2\\
35 &    SDSSJ2303+2028 & 23:03:01.6 & +20:28:20.5 & 6121 &  70 & 56187 & 0.788 & type 2 & 7.32 & 12.40 & 46.26 & 1 &   890.7  $\pm$  467.9 &  857.3  $\pm$  170.9\\
36 &    SDSSJ2351+1248 & 23:51:20.1 & +12:48:19.9 & 6145 & 163 & 56266 & 0.052 & type 2 & 7.11 & 10.52 & 44.21 & 1 &  -100.9  $\pm$   20.5 &   99.6  $\pm$    0.5\\
\enddata
\tablenotetext{a}{The coordinates in the SDSS DR12.}
\tablenotetext{b}{1: type 1 AGN. 2: type 2 AGN (see Section \ref{DA}).}
\tablenotetext{c}{The infrared luminosity at 8--1000 $\micron$ derived in \cite{Toba_16}.}
\tablenotetext{d}{The bolometric luminosity calculated by integrating the best-fit SED template  at wavelengths longward of Ly$\alpha$ (see Section \ref{VVD_IR}).}
\tablenotetext{e}{0: there is no broad wing of \oiii line. 1: there is broad wing of \oiii line (see Section \ref{Spfit}).}
\end{deluxetable*}  
   
\end{document}